\documentclass[prb,showpacs,preprintnumbers,amsfonts,amssymb,amsmath,floats,twocolumn,aps]{revtex4}

\usepackage{graphicx}
\usepackage{dcolumn}
\usepackage{bm}
\usepackage[final,dvips]{epsfig}
\usepackage{bm}
\usepackage{color}
\usepackage{soul}
\usepackage{subfigure}

\newcommand{\ff}[1]{{\boldsymbol #1}}
\newcommand{\ca}[1]{{\cal #1}}

\begin{document} 
  
\title{
Ferromagnetism of magnetic impurities coupled indirectly via conduction electrons:
Insights from various theoretical approaches
}

\author{Irakli Titvinidze, Andrej Schwabe and Michael Potthoff}

\affiliation{I. Institut f\"ur Theoretische Physik, Universit\"at Hamburg, Jungiusstra\ss{}e 9, 20355 Hamburg, Germany}

\begin{abstract}
The magnetic ground-state properties of the periodic Anderson model with a regular depletion of the correlated sites are analyzed within different theoretical approaches.
We consider the model on the one-dimensional chain and on the two-dimensional square lattice with hopping between nearest neighbors.
At half-filling and with correlated impurities present at every second site, the depleted Anderson lattice is the most simple system where the indirect magnetic coupling mediated by the conduction electrons is ferromagnetic.
We discuss the underlying electronic structure and the possible mechanisms that result in ferromagnetic long-range order.
To this end, different numerical and analytical concepts are applied to the depleted Anderson and also to the related depleted Kondo lattice and are contrasted with each other.
This includes numerical approaches, i.e.\ Hartree-Fock theory, density-matrix renormalization and dynamical mean-field theory, as well as analytical concepts, namely a variant of the Lieb-Mattis theorem and the concept of flat-band ferromagnetism, and finally perturbative approaches, i.e.\ the effective RKKY exchange in the limit of weak and the ``inverse indirect magnetic exchange'' in the limit of strong coupling between the conduction band and the impurities.
\end{abstract} 
 
\pacs{71.70.Gm,75.10.Lp,75.75.-c} 


\maketitle 

\section{Introduction}

Local magnetic moments resulting from partially filled localized orbitals can experience an indirect magnetic exchange coupling mediated via a system of conduction electrons. 
The prime example is the Ruderman-Kittel-Kasuya-Yoshida (RKKY) effective interaction \cite{RK54,Kas56,Yos57} which has an oscillatory dependence on the distance between the magnetic impurities. \cite{Yos96,NR09}
In particular, the RKKY mechanism in many cases explains the coupling of magnetic adatoms on non-magnetic metallic surfaces. 
Due to the recent methodical advances in spin-resolved scanning tunneling microscopy techniques \cite{Wie09} it is nowadays possible to map out the strength and the oscillatory distance dependence of the RKKY interaction with atomic resolution. \cite{HLH06,ZWL+10}
Moreover, the possibility to manipulate the positions of individual magnetic atoms offers the exciting perspective to build artificial adatom magnetic nanostructures with tailored magnetic properties. \cite{KWCW11,KWC+12}
In view of possible future applications for magnetic data storage, ferromagnetically ordered nanostructures deserve particular attention.

The simplest theoretical model of correlated electrons on a lattice that may sustain ferromagnetism induced by indirect RKKY coupling is the periodic Anderson model. \cite{And61,LC79}
Here, the magnetic impurities or the magnetic adatoms are described by sites with a non-zero on-site Hubbard-type interaction of strength $U$. 
The impurity sites are coupled via a hybridization term $\propto V$ to a system of non-interacting conduction electrons which hop between the nearest-neighboring sites of a lattice with a hopping amplitude $t$.
Local magnetic moments on the impurity sites are formed in the strong-coupling limit $U \gg V^{2} \rho_{0}$ where $\rho_{0}$ is the conduction-electron density of states at the Fermi energy. 
A local polarization of the conduction electrons is induced via the antiferromagnetic local Kondo exchange \cite{SW66} $J=8V^2/U$ and results in an effective interaction $J_{\rm RKKY, ij} = -J^2 \chi_{ij}^{(0)}$ between the impurities coupling to lattice sites $i$ and $j$ that is given in terms of the non-local static magnetic susceptibility $\chi_{ij}^{(0)}$.

Consider the conceptually most simply case of a half-filled and particle-hole symmetric periodic Anderson model on a bipartite $D$-dimensional lattice with $L$ conduction-electron sites. 
As the conduction-electron spin-spin correlations between nearest-neighbor sites are antiferromagnetic in this case, i.e.\ $\chi_{ij}^{(0)} < 0$, a {\em ferromagnetic} RKKY coupling is only possible if impurities are placed at sites that are separated by two (or generally by an even number) of nearest-neighbor hops.
Therefore, a corresponding spin-depleted periodic Anderson model with at most $R=L/2$ impurity spins is expected to sustain ferromagnetic order.

Apart from the perspectives in modeling artificial ferromagnetic nanostructures, the depleted periodic Anderson as well as the related Kondo model are interesting to describe the electronic structure of one-dimensional organic ferromagnets. \cite{KMOS87,YEL+06,HGWX07,HWR+14} 
The limit of strong Kondo coupling $J$ is accessible to studies of ultracold alkali earth atoms trapped in optical lattices as has been discussed recently. \cite{GHG+10,FFHR10} 

Here, however, we would like to emphasize that the considered depleted models are in first place interesting from a more fundamental point of view.
Due to the geometrically regular removal of impurities, an insulating state is avoided, and {\em metallic} ferromagnetism can be studied in a simple particle-hole symmetric model. 
This is a similar motivation as for the antiferromagnetic case studied in Ref.\ \onlinecite{Ass02}.

Another motivation for the present study is that there are several and conceptually very different theoretical approaches that apply to the case of the depleted Anderson or Kondo lattice and provide independent explanations for the emergence of ferromagnetic order: 
(i) First of all, beyond the regime of weak hybridization $V$, the indirect exchange is expected to compete \cite{Don77} with the Kondo screening \cite{Hew93} of the impurity magnetic moments. 
(ii) Interestingly, in the limit of strongly coupled impurities (strong $V$ or strong $J$), the Kondo effect has actually been seen to {\em corroborate} ferromagnetic ordering. 
Namely, an ``inverse'' indirect magnetic exchange (IIME) provides a coupling of local moments formed {\em between} almost localized Kondo singlets. \cite{STP13}
(iii) Moreover, a variant \cite{She96,Tsu97a} of the Lieb-Mattis theorem \cite{LM62a,Lie89} may apply which predicts a ferro- or actually ferrimagnetic ground state in the entire $V$ or $J$ regime. 
This is compatible with a crossover from the RKKY to the IIME limit.
(iv) Even the non-interacting ($U=0$) limit is interesting: 
The lattice geometry inevitably leads a completely non-dispersive band right at the Fermi energy.
As a result there is a highly degenerate Fermi sea where the configuration with a fully polarized flat band is one of the possible many-body ground states.
This raises the question which characteristics of the interacting ferromagnet already derive from those of the fully polarized non-interacting ground state.
Moreover, the Stoner criterion \cite{Sto51} applies to the weakly interacting system, i.e.\ the flat band makes the system extremely susceptible to magnetic ordering.
(v) Another question is whether the ferromagnetic order can be understood within the concept of ``flat-band ferromagnetism'' \cite{Mie91a,Tas92,MT93,Tas97} as well. 
Comparison of numerically exact data with standard Hartree-Fock theory may show whether the ground state is correlated at all and thus different from the fully polarized Fermi sea that is predicted in the flat-band ferromagnetism scenario.

Finally, another motivation of the present paper is a methodical one:
The physics of the models studied here are very well accessible to reliable numerical techniques. 
The density-matrix renormalization group (DMRG) \cite{Whi92,Sch11} provides essentially exact results for the one-dimensional case. 
Interestingly, dynamical mean-field theory (DMFT) \cite{MV89,GKKR96} turns out as very reliable, too, even for $D=1$, since the correlated sites are separated by 4 or more nearest-neighbor hops and thus the local approximation for the self-energy becomes rather accurate.
This allows us to address the two-dimensional lattice as well and motivates us to analyze the low-energy part of the single-particle excitation spectrum to study the ``fate'' of the flat band for a strongly correlated system. 
Results will be discussed mainly for the case of Anderson but also for Kondo impurities.

The paper is organized as follows:
In the next section, we introduce the depleted Anderson model. 
Its properties in the non-interacting limit and in particular the emergence of a flat band are analyzed in Sec.\ III. 
A study of the properties of the non-interacting state with a fully polarized flat band is given in Sec.\ IV.
The related concept of flat-band ferromagnetism is discussed in Sec.\ V along with calculations based on Hartree-Fock theory for the strong-coupling (strong $U$) regime.
DMRG data for the one-dimensional model in the whole $V$ range are presented in Sec.\ VI.
This not only demonstrates the limitations of the Hartree-Fock approach but also allows us to analyze the crossover from the weak-$V$ to the strong-$V$ limit. 
The role of the Lieb-Mattis theorem in this context is clarified in Sec.\ VII, while in Sec.\ VIII the underlying physical pictures and mechanisms are discussed.
Sec.\ IX introduces the dynamical mean-field approach and therewith allows us to shift the perspective and to consider two-dimensional systems and particularly the low-energy one-particle excitation spectrum.
This is worked out in Secs.\ X and XI for the coherent part of the Green's function and the density of states, respectively. 
In Sec.\ XII once more shifts and extends the perspective by considering the depleted Kondo model in a regime where it is different from the related Anderson model. 
Qualitatively, the low-energy excitation spectra of the two models turn out as very similar as discussed in Sec.\ XIII along with the question for the fate of the flat band in the strongly interacting limit.
The smaller local Hilbert space of the depleted Kondo model allows to carry out precise DMRG calculations for the charge susceptibility, discussed in Sec.\ XIV, which underpin the interpretations of the dynamical mean-field analysis. 
A summary of the main results and the conclusions are given in Sec.\ XV.

\section{Depleted periodic Anderson model}

\begin{figure}[t]
\centerline{\includegraphics[width=0.4\textwidth]{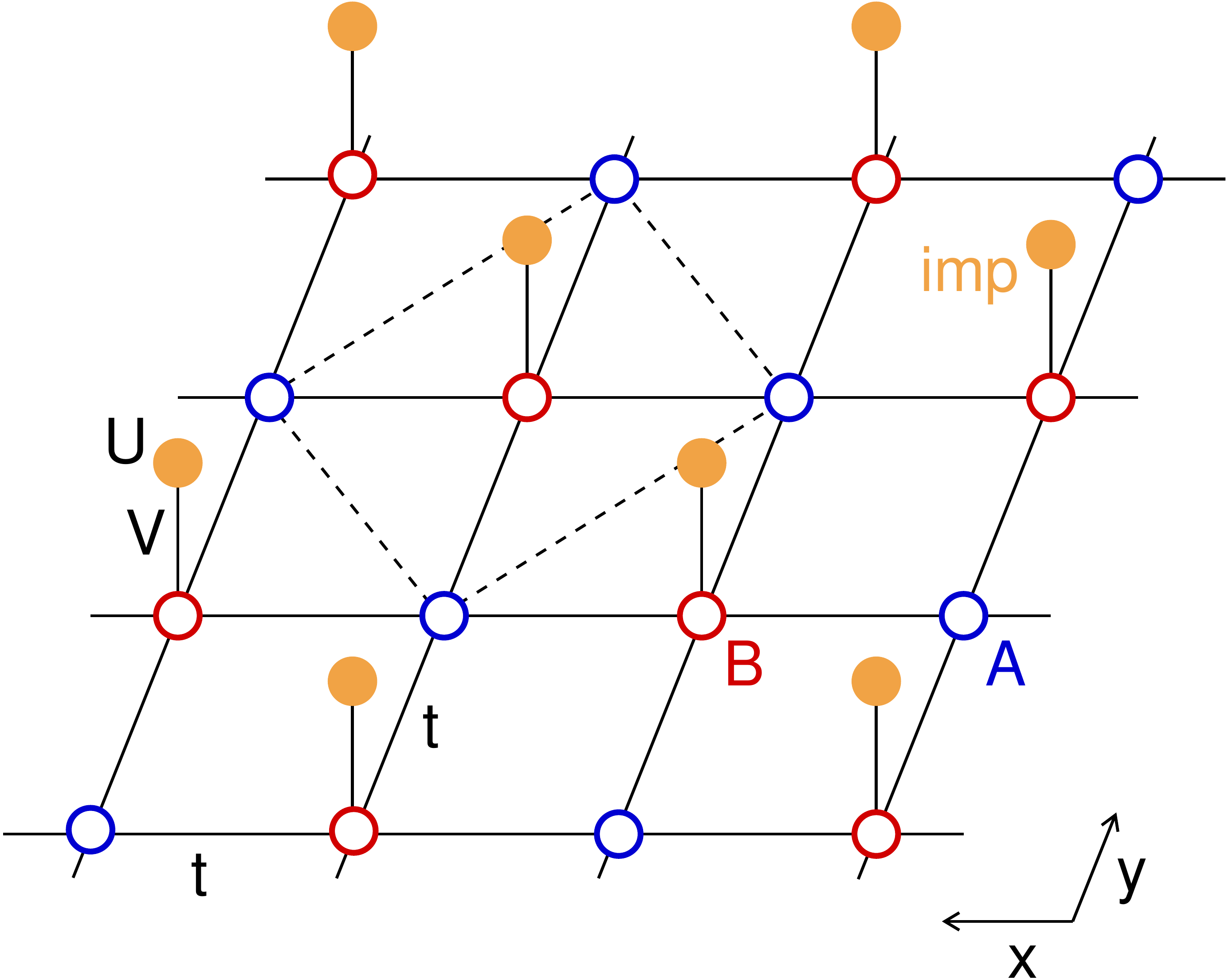}} 
\caption{
(Color online) 
Geometry of the lattice model considered. 
$R$ impurities (orange) with on-site Hubbard interaction $U$ are coupled via a hybridization of strength $V$ to the B sites (red) of a bipartite lattice with $L=2R$ sites in total (A sites: blue). 
Periodic boundary conditions are assumed.
The energy scale is fixed by setting $t=1$ for the nearest-neighbor hopping in the conduction-band system. 
Dashed lines indicate a unit cell. 
The system is considered at half-filling, i.e., the average total particle number is $\langle N \rangle = L+R$.
}
\label{fig:geo}
\end{figure}

We consider conduction electrons hopping between the nearest-neighbor sites of a $D$-dimensional bipartite lattice ($D=1$ chain and $D=2$ square lattice) consisting of $L$ sites.
The hopping amplitude fixes the energy scale, i.e.\ $t=1$.
``Impurities'', i.e.\ sites with a finite local Hubbard interaction $U$, are coupled via a hybridization of strength $V$ to the B sites of the lattice consisting of the two sublattices A and B, see Fig.\ \ref{fig:geo} for the two-dimensional case.
In total, there are $R=L/2$ impurities.
The Hamiltonian reads as:
\begin{eqnarray}
{\cal H}
&=&
-t \sum_{\langle ij \rangle ,\sigma} \left( a^\dagger_{i\sigma} b_{j\sigma} + \mbox{h.c.} \right)
+
V \sum_{j \in B, \sigma}\left(b^\dagger_{j\sigma}c^{\phantom\dagger}_{k_{j}\sigma} + \mbox{h.c.} \right) 
\nonumber \\
&+&
U\sum_{j \in B}n_{k_{j} \uparrow}^{(c)} n_{k_{j}\downarrow}^{(c)}
- 
\mu \sum_{i \in A,\sigma} n_{i\sigma}^{(a)} 
-
\mu
\sum_{j \in B,\sigma} n_{j\sigma}^{(b)} 
\nonumber \\
&+&
(\varepsilon - \mu) \sum_{j \in B,\sigma} n_{k_{j}\sigma}^{(c)}
\, .
\label{eq:ham} 
\end{eqnarray}
Here $a_{i\sigma}^\dagger$, $b_{j\sigma}^\dagger$ and $c_{k\sigma}^\dagger$ create an electron with spin $\sigma=\uparrow,\downarrow$ at a site $i$ in sublattice A, at a site $j$ in sublattice B and at an impurity site $k$, respectively. 
With $k_{j}$ we denote the impurity site attached to the B site $j$. 
Furthermore,
$n_{i\sigma}^{(a)}=a^\dagger_{i\sigma}a^{\phantom\dagger}_{i\sigma}$, 
$n_{j\sigma}^{(b)}=b^\dagger_{j\sigma}b^{\phantom\dagger}_{j\sigma}$ and 
$n_{k\sigma}^{(c)}=c^\dagger_{k\sigma}c^{\phantom\dagger}_{k\sigma}$ denote the corresponding occupation-number operators. 
We will consider the model at half-filling where the average total particle number $\langle N \rangle = L+R$. 
This is ensured by choosing $\mu=0$ for the chemical potential and $\varepsilon=-U/2$ for the on-site energy of the impurity sites.

\section{Non-interacting limit} 
\label{sec:non}

Assuming periodic boundary conditions and exploiting translational symmetries, the non-interacting part of the Hamiltonian is block-diagonalized by switching to a wave-vector representation. 
There are 3 sites in a unit cell (A, B, impurity). 
Consequently, 
\begin{equation}
\label{H0} 
{\cal H}_0=\sum_{{\bf k}\sigma} \left(a^\dagger_{{\bf k}\sigma},b^\dagger_{{\bf k}\sigma}, c^\dagger_{{\bf k}\sigma}\right)
\left(\hat t ({\bf k}) -\mu \right) \left(
\begin{array}{c}
a_{{\bf k}\sigma}^{\phantom\dagger}\\
b_{{\bf k}\sigma}^{\phantom\dagger}\\ 
c_{{\bf k}\sigma}^{\phantom\dagger}
\end{array}
\right) \: 
\end{equation}
with a $3\times 3$ hopping matrix for each wave vector $\ff k$:
\begin{equation}
\label{hopping}
\hat t ({\bf k}) = \left(
\begin{array}{ccc}
0 & \varepsilon({\bf k}) & 0 \\
\varepsilon({\bf k}) & 0 & V\\
0 &V & 0
\end{array}
\right) \: .
\end{equation}
Here, 
\begin{equation}
\varepsilon({\bf k})=-2t \sum_{s=1}^D \cos k_s  
\end{equation}
is the bare ($V=0$) dispersion of the conduction-electron band. 
Diagonalization of ${\cal H}_0$,
\begin{equation}
\label{dispersion}
{\eta}_{m}({\bf k})
=
\left[ \hat Q^\dagger_0({\bf k}) \hat t ({\bf k}) \hat Q_0({\bf k}) \right]_{mm}  \; ,
\end{equation}
is achieved with the unitary transformation matrix
\begin{equation}
\hat Q_{0}({\bf k})= \frac{1}{\sqrt{2}} \frac{1}{\xi(\ff k)}
\left(
\begin{array}{ccc}
\varepsilon({\bf k}) & \sqrt{2} V & \varepsilon({\bf k}) \\
-\xi(\ff k)  & 0 & \xi(\ff k) \\
V & - \sqrt{2} \varepsilon({\bf k}) & V
\end{array}
\right) \: .
\label{eq:qmat}
\end{equation}
This yields three bands ($m=1,2,3$), 
\begin{eqnarray}
\label{eq:spectrum} 
\eta_{1}({\bf k}) &=& - \xi({\bf k}) - \mu  
\nonumber \\
\eta_{2}({\bf k}) &=&  - \mu  
\nonumber \\
\eta_{3}({\bf k}) &=& \xi({\bf k}) - \mu  
\end{eqnarray}
where $\xi({\bf k})=\sqrt{\varepsilon({\bf k})^2+V^2}$. 
Each band consists of $L/2$ eigenenergies, i.e.\ there are $L/2$ allowed wave vectors in the first Brillouin zone.
The $m=2$ band is dispersionless, i.e.\ ``flat''.


The presence of the flat band originates from the (i) bipartite lattice structure, (ii) the restriction to inter-sublattice hopping, (iii) homogeneous on-site energies and (iv) from the fact that there are more A sites (including impurity sites) than B sites
(for the present discussion we have to regard the impurity sites as A sites as they are coupled to B sites only).
This can easily be seen from the real-space representation of the hopping matrix $\hat t$. 
We have
\begin{eqnarray}
t_{ii'} &=& \varepsilon \delta_{ii'} \quad \mbox{for} \; i,i' \in A
\nonumber \\
t_{jj'} &=& \varepsilon \delta_{jj'} \quad \mbox{for} \; j,j' \in B
\nonumber \\
t_{ij} &=& t_{ji}^{\ast} = T_{ij} \quad \mbox{for} \; i \in A, j\in B 
\: ,
\end{eqnarray}
where $\varepsilon$ is an arbitrary real parameter and $\hat T$ an arbitrary matrix of dimensions $L_{\rm A} \times L_{\rm B}$ where $L_{\rm A}$ is the number of A-sites and $L_{\rm B}$ the number of B sites (in our case, $L_{\rm A} = L/2 + R = L$, $L_{\rm B} = L/2$).
We choose $\varepsilon=0$ for simplicity. 
An eigenvector $u$ of $\hat t$ corresponding to the eigenvalue $\eta=0$ must satisfy
\begin{eqnarray}
\sum_{j\in B} T_{ij} u_{j} = 0 \qquad \forall \; i \in A \; ,
\nonumber \\
\sum_{i\in A} T_{ji}^{\dagger} u_{i} = 0 \qquad \forall \; j \in B
\: .
\end{eqnarray}
$u$ has $L_{\rm A}$ components $u_{i}$ referring to A sites and $L_{\rm B}$ components $u_{j}$ referring to B sites.
The first set of $L_{\rm A}$ conditional equations is satisfied by setting $u_{j}=0$ for all $j\in B$.
The second set of equations represents a homogenous set of $L_{\rm B}$ linear equations for $L_{\rm A}$ unknowns.
Consequently, there are at least $L_{\rm A} - L_{\rm B}$ linear independent and normalized solutions, i.e.\ the eigenvalue $\eta=0$ is at least $L_{\rm A}-L_{\rm B}$-fold degenerate (in our case, the degeneracy is $L/2$).

\section{Fully polarized state} 

For $\mu=0$ and in the non-interacting case for $U=0$, the $m=1$ band is fully occupied by $L$ electrons, and the $m=3$ band is completely empty. 
Further $L/2$ electrons with spin $\sigma=\uparrow,\downarrow$ occupy the flat band $m=2$. 
This leads to a $2^{L/2}$-fold degeneracy of the non-interacting ground-state energy which is expected to be lifted due to interactions.

The Stoner criterion \cite{Sto51} shows that the paramagnetic ground state is unstable towards ferromagnetic ordering at arbitrarily small but finite $U>0$. 
Let us define the ground-state expectation value of the ordered magnetic moment on A, B and impurity sites as
\begin{eqnarray}
m_{\rm A} &=& \langle n_{i\uparrow}^{(a)} \rangle - \langle n_{i\downarrow}^{(a)} \rangle \: ,
\nonumber \\
m_{\rm B} &=& \langle n_{j\uparrow}^{(b)} \rangle - \langle n_{j\downarrow}^{(b)} \rangle \: ,
\nonumber \\
m_{\rm imp} &=& \langle n_{k\uparrow}^{(c)} \rangle - \langle n_{k\downarrow}^{(c)} \rangle \: .
\label{eq:magdef}
\end{eqnarray}
That the ground state, for weak $U>0$, is in fact fully polarized, i.e.\ that $m = m_{\rm A} + m_{\rm B} + m_{\rm imp} = \pm 1$ can be seen easily by first-order perturbation theory in $U$: 
As is detailed in Appendix \ref{appendix:1OC}, the fully polarized state has lowest total energy.

At $U=0$ we therefore pick the fully polarized state with total magnetization $m=1$ and study the $V$-dependence of the different contributions to $m$. 
This can be done by computing the spin-dependent occupation numbers for the sites $\alpha={\rm A,B,imp}$ in a unit cell via
\begin{eqnarray}
  \langle n_{\sigma}^{(\alpha)} \rangle
  = 
  \frac{2}{L} \sum_{\bf k} \sum_{m=1}^3 \left| Q_{0,\alpha m}({\bf k}) \right|^2 
  f_{\ff k, m,\sigma} \: ,
\label{eq:ns}  
\end{eqnarray}
where $f_{\ff k, m=1,\sigma} = 1$ for the lowest band, $f_{\ff k, m=3,\sigma} = 0$ for the highest band, whereas
$f_{\ff k, m=2,\uparrow} = 1$ and $f_{\ff k, m=2,\downarrow} = 0$ for the flat band.
Using Eqs.\ (\ref{eq:qmat}) and (\ref{eq:magdef}), this yields the following site-dependent magnetic polarization in the $U\to 0$ limit: 
\begin{eqnarray}
  m_{\rm A}  
  &=& 
  \frac{2}{L} \sum_{\bf k} \frac{V^2}{\varepsilon({\bf k})^2+V^2} \: ,
\nonumber \\
  m_{\rm B}
  &=&
  0 \: ,
\nonumber \\
  m_{\rm imp}
  &=&
  \frac{2}{L} \sum_{\bf k} \frac{\varepsilon({\bf k})^2}{\varepsilon({\bf k})^2+V^2} \: .
\label{eq:ms}  
\end{eqnarray}
In the thermodynamical limit $L\to \infty$, the $\ff k$-sums can be rewritten as an energy integration weighted by the free tight-binding density-of-states and evaluated numerically for $D=2$. 
The result is shown in the upper part of Fig.\ \ref{fig:mag} (dashed lines).
For $D=1$ the calculation can be done analytically. 
We find $m_{\rm B}=0$ and
\begin{equation}
  m_{\rm A} = \frac{V}{\sqrt{V^2+4t^2}} \: , \quad
  m_{\rm imp} = \frac{\sqrt{V^2+4t^2}-V}{\sqrt{V^2+4t^2}} \: .
\end{equation}
These are plotted in the lower part of Fig.\ \ref{fig:mag} (dashed lines).

As can be seen in the figure, there is a crossover of the fully polarized magnetic state of the non-interacting system from the weak-$V$ to the strong-$V$ regime. 
This crossover is in some respects reminiscent of the corresponding crossover of the strongly interacting system for large $U$, namely from a ferromagnet driven by the effective indirect RKKY interaction for weak $V$ to a ferromagnet driven by the effective inverse indirect exchange interaction (IIME) for strong $V$ (see discussion in Secs.\ \ref{sec:dmrg} and \ref{sec:iime}).

For $V=0$, electrons at impurity sites are perfectly localized and form a local spin $1/2$.
In the fully polarized state, the resulting magnetic moment is $m_{\rm imp}=1$.
For finite but weak $V$, electrons delocalize. 
In the non-interacting system there is only the Fermi-gas contribution to the local moments, and sizeable moments are only formed in the symmetry-broken state considered here.
These ordered local moments are mainly present on the impurity sites. 
In the one-dimensional case we have $m_{\rm imp} \simeq 1 - V/2t \to 1$ and $m_{\rm A} \simeq V/2t \to 0$ for $V\to 0$, 
while for $D=2$ there is a non-analytical behavior of the moments $m_{\rm imp} \to 1$ and $m_{\rm A} \to 0$.
On the other hand, in the limit $V \to \infty$, electrons are perfectly localized at A sites, and thus $m_{\rm A} = 1$ in the symmetry-broken state.
For finite but strong $V$, delocalization of A-site electrons then tends to reduce $m_{\rm A}$. 
We have $m_{\rm A} \simeq 1- qt^{2}/V^{2}$ and $m_{\rm imp} \simeq qt^{2}/V^{2}$, where $q=2D$ is the coordination number.
In the entire $V$ range, and also independent of the dimensionality, the magnetic polarization at the B-sites remains zero, and hence $m_{\rm A} + m_{\rm imp} = 1$.

\section{Hartree-Fock theory vs.\ flat-band ferromagnetism}

Let $| {\rm F0} \rangle$ be the Fermi sea that is constructed by completely filling the $m=1$ band with $L$ electrons and the $m=2$ band with $L/2$ spin-$\uparrow$ electrons, i.e.\ $| {\rm F0} \rangle$ is a single Slater determinant with a fully polarized $m=2$ band:
\begin{equation}
  | {\rm F0} \rangle = \prod_{\ff k} c^{\dagger}_{\ff k, m=2, \uparrow} \prod_{\ff k, \sigma} c^{\dagger}_{\ff k,m=1,\sigma} | {\rm vacuum} \rangle \: .
\end{equation}
For $U=0$, this is only one ground state among the $2^{L/2}$ ground states.
Obviously, the macroscopically large degeneracy is due to the fact that the $m=2$ band is flat.
``Flat-band ferromagnetism'' was recognized by Mielke and Tasaki \cite{Mie91a,Mie91b,Mie92,Tas92,MT93,Tas97} as a possible route to itinerant ferromagnetism in the Hubbard model.
For the Hubbard model on lattices with certain topologies (not necessarily bipartite) and at certain fillings (not necessarily half-filling), a fully polarized ferromagnetic state can be obtained by filling the flat band with $\uparrow$ electrons only.
The very non-trivial and exact statement made by Mielke and Tasaki is that this state becomes the {\em unique} ground state for any $U>0$ (apart from the $(2S_{\rm tot}+1)$-fold degeneracy).
While the fully polarized state is not affected by $U$, all other ground states gain a higher energy. 

The situation considered here is somewhat different because the Hubbard interaction is not present at all sites and also because there is a dispersive band below the flat band in the non-interacting band structure (see, however, Ref.\ \onlinecite{Tas95}). 
In fact, we can easily show that $| {\rm F0} \rangle$ cannot be the exact ground state for any $U>0$ and that, therefore, the concept of flat-band ferromagnetism does not apply to the present case.

The argument is based on Hartree-Fock theory which can be defined by the mean-field decoupling
\begin{equation}
  n_{k\uparrow}^{(c)} n_{k\downarrow}^{(c)} \mapsto \langle n_{k\uparrow}^{(c)} \rangle n_{k\downarrow}^{(c)}+ 
  n_{k\uparrow}^{(c)} \langle n_{k\downarrow}^{(c)} \rangle -\langle n_{k\uparrow}^{(c)}\rangle\langle n_{k\downarrow}^{(c)} \rangle 
\label{MF_decoupling}
\end{equation}
which replaces the original Hamiltonian Eq.\ (\ref{eq:ham}) by a mean-field Hamiltonian that is bilinear in creators and annihilators. 
The solution of the resulting effective Hartree-Fock equations must be obtained by computing the expectation values self-consistently from the mean-field Hamiltonian.
At zero temperature, the approach is fully equivalent with the variational optimization of a Slater determinant consisting of $N$ {\em a priori} unknown single-particle orbitals using the Ritz principle. 
Note that this {\em ansatz} also comprises the Slater determinant $| {\rm F0} \rangle$ in particular. 

The numerical calculations clearly show that the optimized Hartree-Fock ground state $| {\rm HF} \rangle$ has a lower total energy:
$\langle {\rm HF} | \ca H | {\rm HF} \rangle < \langle {\rm F0} | \ca H | {\rm F0} \rangle$ if $V\ne 0$ and $V\ne \infty$.
Qualitatively, this can be understood in the following way:
For any $V$ with $0<V<\infty$, the flat band is fully polarized and singly occupied with $\uparrow$-electrons in the state $| {\rm F0} \rangle$.
Nevertheless, there is always a finite double occupancy on the impurity sites, $\langle {\rm F0} | n^{{(c)}}_{\uparrow} n^{{(c)}}_{\downarrow} | {\rm F0} \rangle > 0$ as the impurity orbitals have contributions from all three bands 
(already this fact is conflicting with flat-band ferromagnetism, see Ref.\ \onlinecite{Tas97}, for example). 
Hartree-Fock theory tends to reduce this double occupancy,
$\langle {\rm HF} | n^{{(c)}}_{\uparrow} n^{{(c)}}_{\downarrow} | {\rm HF} \rangle <
\langle {\rm F0} | n^{{(c)}}_{\uparrow} n^{{(c)}}_{\downarrow} | {\rm F0} \rangle$,
by increasing the polarization at the impurity sites, $m_{\rm imp}^{{\rm (HF)}} > m_{\rm imp}^{{\rm (F0)}}$.
Thereby, the interaction energy is lowered. 

\begin{figure}[t]
\centerline{\includegraphics[width=0.4\textwidth]{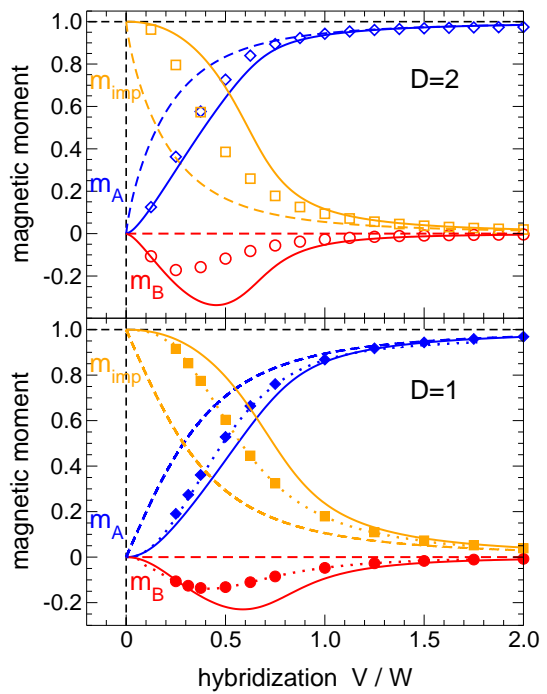}} 
\caption{
(Color online) 
Ordered magnetic moments on the impurity sites, $m_{\rm imp}$, and on A and B sites, $m_{\rm A}$ and $m_{\rm B}$, of the $D=2$ (top) and the $D=1$ (bottom) lattice as functions of the hybridization strength $V/W$ where $W=8$ ($D=2$) and $W=4$ ($D=1$) is the free band width.
Dashed lines: Fully polarized ground state of the non-interacting system ($U=0$).
Solid lines: Results of a Hartree-Fock calculation for $U=2W$.
Filled symbols for $D=1$: DMRG calculation \cite{STP13} for $U=2W$ ($L=49$ lattice sites, $R=25$, open boundary conditions, results are shown for the central unit cell).
Dotted line for $D=1$: DMFT calculation \cite{STP13} for $U=2W$ ($L=100$ lattice sites).
Open symbols for $D=2$: DMFT calculation for $U=2W$ ($L=1000\times 1000$).
For the DMFT calculations, periodic boundary conditions are assumed.
}
\label{fig:mag}
\end{figure}

The magnetic polarizations at the different sites, as obtained from Hartree-Fock theory, are shown by the solid lines in Fig.\ \ref{fig:mag}. 
Calculations have been done for $U=8$ ($D=1$) and $U=16$ ($D=2$).
One clearly notes that (i) $m_{\rm imp}^{{\rm (HF)}}$ is different from $m_{\rm imp}^{{\rm (F0)}}$ for all $V$ with $0<V<\infty$, which already reflects that $|{\rm F0} \rangle$ cannot be the ground state, and that (ii) $m_{\rm imp}^{{\rm (HF)}} > m_{\rm imp}^{{\rm (F0)}}$ due to the reduced double occupancy on the impurity sites.
Opposed to $|{\rm F0} \rangle$, one finds a non-zero but negative polarization at the B sites.
This is an indication of local antiferromagnetic Kondo correlations.
The total magnetization is still unity, $m=m_{\rm imp} + m_{\rm A} + m_{\rm B}=1$.

\section{Density-matrix renormalization group}
\label{sec:dmrg}

For the one-dimensional case, the Hartree-Fock results can be compared with numerically exact data from Ref.\ \onlinecite{STP13} obtained by the density-matrix renormaliization group. \cite{Whi92,Sch11}
DMRG calculations have been performed with a standard implementation using matrix-product forms for the trial many-body state as well as for the operators (see Ref.\ \onlinecite{TSRP12} for some details). 
The data shown in Fig.\ \ref{fig:mag}b (filled symbols) have been obtained for a system with $L=49$ sites and open boundaries. 

DMRG indeed predicts a ferromagnetic ground state with a total magnetization independent of $V$ (see Secs.\ \ref{sec:lm} and \ref{sec:iime} for further discussion). 
The magnetic polarizations at the different sites have been plotted for the central unit cell of the chain for the sector with maximum magnetic quantum number $M_{\rm tot} = S_{\rm tot}$.
The DMRG ground state is unique, apart from the $2S_{\rm tot}+1$-fold degeneracy.
On the scale used in Fig.\ \ref{fig:mag}, the DMRG results are representative for an infinite chain as has been verified by comparing data for different system sizes. 

For $V\to 0$ and for $V\to \infty$, Hartree-Fock theory reproduces the exact result for $m_{\rm A}$, $m_{\rm B}$ $m_{\rm imp}$ and for the total magnetization $m$. 
For intermediate $V$, however, there are strong deviations: 
Here, Hartree-Fock theory overestimates the absolute strengths of the polarizations on the impurity and on the B sites, and underestimates $m_{A}$.

We conclude that Hartree-Fock theory is able to roughly describe the trends of the magnetic polarizations with $V$ but does not allow for quantitative predictions and even qualitatively fails to describe the magnetic order in terms of spin-correlation functions. 
The magnetic ground state of the considered periodic Anderson model with depleted magnetic impurities must therefore be seen as being correlated. 
It cannot be described by a fully polarized Fermi sea or by any other (variationally optimized) Slater determinant.

\section{Lieb-Mattis theorem}
\label{sec:lm}

Although the ground state is correlated, there are exact statements available for the particular geometrical and electronic structure considered here. 
This goes back to the Lieb-Mattis theorem \cite{LM62a} which states that the antiferromagnetic spin-$1/2$ Heisenberg model on an arbitrary bipartite lattice has a unique singlet ground state if the number $L_{\rm A}$ of A sites equals the number $L_{\rm B}$ of B sites.
For $L_{\rm A}>L_{\rm B}$, the ground state has total spin $S_{\rm tot} = (L_{\rm A} - L_{\rm B})/2$ and is unique apart from the trivial $(2S_{\rm tot} +1)$-fold degeneracy.
The proof of the theorem is based on the positivity of the Hamiltonian after a unitary spin-reflection transformation which permits the application of the Perron-Frobenius theorem.
The concept of ``spin-reflection positivity'' has been used to generalize the Lieb-Mattis theorem to the half-filled repulsive Hubbard model with constant $U_{i}=U>0$ for any site $i$ on a bipartite lattice. \cite{Lie89}
For the half-filled and symmetric periodic Anderson model (with as many localized as conduction orbitals, i.e., $R=L$) on a bipartite lattice, a modified proof is possible. \cite{UTS92,TSU97b}
Analogous results are also available for the half-filled Kondo-lattice model. \cite{She96,Tsu97a}

Let us now focus on the depleted periodic Anderson model at half-filling that is considered here.
As there are additional sites (the A sites) with vanishing Hubbard-type interaction, the ground state is not necessarily unique, i.e.\ the Lieb-Mattis theorem (and generalizations thereof) cannot be applied in this respect.
In fact, for certain $L$ and $R$, one can easily find examples with a degenerate ground state. 
The most simple case is probably given by $L=3$ and $R=1$ where the impurity site is attached to the central site of the three-site tight-binding chain with open boundaries. 
A simple calculation shows that the eigenstates of lowest energy are given by a triplet, $S_{\rm tot} = 1 = (L_{\rm A} - L_{\rm B})/2$, and a singlet, $S_{\rm tot}=0$, for any $U>0$, i.e.\ the ground-state energy is four-fold degenerate.
However, one of the ground states (if there is degeneracy) is in the sector with $S_{\rm tot} = 1 = (L_{\rm A} - L_{\rm B})/2$. 

This is similar to the depleted Kondo lattice.
There is a theorem, proven by Shen, \cite{She96} stating that for $R$ localized spins $1/2$ coupling via a local antiferromagnetic exchange to a half-filled system of otherwise non-interacting conduction electrons on a bipartite $D$-dimensional lattice with $L\ge R$ sites, there is at least one ground state with $S_{\rm tot} = (L_{\rm A} - L_{\rm B})/2$. 
Degeneracy can be excluded for the dense case, $R=L$, or for a finite Hubbard-type interaction $U>0$ among the conduction electrons.\cite{She96,Tsu97a}

The DMRG calculations (Fig.\ \ref{fig:mag}) for $L=49$ and $R=25$ yield a ground-state total spin $S_{\rm tot} = 12$.
This is consistent with the above prediction, i.e.\ $S_{\rm tot} = (L_{A} - L_{B}) / 2 = (R-1)/2$.
Actually,  the system is ferrimagnetic, also  consistent with the sign of the spin-spin correlation functions that can be predicted exactly. \cite{She96}
Furthermore, the ground state turns out to be unique (apart from the trivial $(2S_{\rm tot}+1)$-fold degeneracy).

\section{Physical mechanisms}
\label{sec:iime}

Opposed to the exact theorems available and opposed to the essentially exact numerical data, perturbative approaches are able to provide a more physical understanding for the emergence of ferromagnetic order.
Moreover, they can explain which sites, depending on $V$, carry the main part of the total magnetic moment and also why the ground state is in fact unique. 

For $U\gg t$ and weak $V$, the model maps onto a depleted Kondo lattice where the correlated sites are replaced by spins $1/2$ coupled via an antiferromagnetic local exchange $J=8V^{2}/U$ to the conduction-electron system. \cite{SW66,SN02} 
As $J$ is small in this limit, RKKY second-order perturbation theory applies. \cite{Yos96,NR09}
At half-filling the effective non-local spin-spin interaction $J_{\rm RKKY} \propto (-1)^{d}$ has an oscillatory dependence on the distance $d$ between two local spins and is ferromagnetic for the model considered here.
Hence, the ferromagnetic ground state results from ferromagnetic RKKY coupling of well-formed local moments at the impurity sites. 
However, the total spin is $S_{\rm tot} = (R-1)/2$ rather than $S_{\rm tot} = R/2$ as could have been expected by simply coupling all impurity spins.

The reason for the ``missing spin 1/2'' is a variant of the Kondo effect that has been studied in Ref.\ \onlinecite{SGP12} and that applies to systems of finite size where perturbation theory in $J$ is well behaved.
Namely, in the weak-coupling limit $V\to 0$ and for the particular system with $L=49$ and $R=25$, exactly one of the impurity spins is Kondo screened, 
and only the remaining ones are subjected to the ferromagnetic RKKY interaction. 

The ``missing spin'' can be understood with the help of the {\em original} Lieb-Mattis theorem:
Using standard degenerate perturbation theory in first order in $J$ for the depleted Kondo lattice, we obtain the following effective model:
\begin{equation}
  H_{\rm eff} = \sum_{j \in B} J_{k_{j}}  \ff S_{k_{j}}  \ff s_{\rm F} \; ,
\label{eq:csm}
\end{equation}
where $\ff S_{k_{j}}$ is a spin $1/2$ at the impurity site $k_{j}$, and $\ff s_{\rm F}$ is the spin of the conduction electron at the Fermi energy.
The effective exchange couplings $J_{k_{j}} = J | \varphi^{(\rm F)}_{k_{j} }|^{2}$ are given in terms of its single-particle wave function $\varphi^{(\rm F)}_{k_{j} }$ and are all positive. 
Eq.\ (\ref{eq:csm}) represents an antiferromagnetic central-spin Heisenberg model.
As this has the topology of a bipartite lattice, the Lieb-Mattis theorem \cite{LM62a} immediately predicts a unique ground state with $S_{\rm tot} = (R-1)/2$.

In the opposite limit of strong hybridization $V\to \infty$, the electrons at the impurity and at the B sites form strongly localized and magnetically inert singlet bound states. 
This implies that the total magnetic moment is no longer carried by the impurity sites. 
However, these ``Anderson singlets'' tend to localize the electrons at the A sites, and this results in the formation of strong local A-site magnetic moments.
There is an indirect coupling between the A-site moments which is ferromagnetic and induced via virtual excitations of the Anderson singlets. 
This is similar to the strong-$J$ limit of the depleted Kondo lattice studied in Ref.\ \onlinecite{STP13}. 
A analogous construction of an effective low-energy model by employing strong-coupling perturbation theory is also possible in the Anderson case (and will be published elsewhere).
The resulting ``inverse indirect magnetic exchange'' (IIME) represents the strong-coupling (strong $V$) analogue of the RKKY mechanism at weak coupling.

The transition from the RKKY to the IIME regime as a function of $V$ is a smooth crossover:
As can be seen in Fig.\ \ref{fig:mag}, the polarization at the impurity sites continuously decreases while the A-site polarization increases.
The negative polarization at the B sites attains a maximal absolute value in the crossover regime $V \sim W/2$. 
The symmetry of the ground state and the total spin $S_{\rm tot}$ does not change with $V$.
In the strong-coupling limit, the total spin $S_{\rm tot} = (R-1)/2$ is due to the ferromagnetic IIME coupling of the $R-1$ A-site moments. 

\section{Dynamical mean-field theory}

It is instructive to compare the results for the magnetic polarization at the different sites that have been obtained by static mean-field theory and (for $D=1$) by DMRG with corresponding results of dynamical mean-field theory (DMFT). \cite{MV89,GKKR96}
The DMFT treats the local correlations exactly, and in particular the Kondo screening of the magnetic impurities. 
It also accurately predicts the indirect inter-impurity magnetic coupling as has been studied before in Ref.\ \onlinecite{TSRP12}.
On the other hand, the feedback of the non-local magnetic correlations on the one-particle excitation spectrum is neglected within DMFT as this would imply non-local contributions to the electron self-energy. 
This feedback effect, however, can be expected as weak for the present case of a system with regularly depleted impurities: 
As the self-energy is non-zero on the correlated sites only, a non-local self-energy diagram must include non-local Green's functions between sites that are separated by 4 or more nearest-neighbor hopping steps. 
An almost local self-energy and thus a single-site DMFT approach should therefore be a reasonable approximation.

We employ a standard implementation of the DMFT using the exact-diagonalization solver, \cite{CK94} i.e.\ the ground state and the single-electron excitation spectrum of the effective impurity problem is obtained by the Lanczos method. 
Calculations have been performed for Anderson impurity models with $n_{\rm s} = 10$ sites. 
A fictitious temperature of $T^{\ast} = 0.002$ is used for the low-energy cutoff of the DMFT self-consistency equation. 
Lattices with periodic boundary conditions and a sufficiently large number of sites $L$ are considered to ensure that the results are free of finite-size errors.
For $U=2W$ and in the entire range of hybridization strengths $V$, there is a ferromagnetic solution of the DMFT equations which is found by running through the usual DMFT self-consistency cycle. 
A paramagnetic solution, on the other hand, could not be stabilized without enforcing spin-symmetric parameters.
We have checked that the ferromagnetic state has the lower total energy. 

The results for the site-dependent polarizations of the depleted Anderson lattice are shown Fig.\ \ref{fig:mag} as dotted lines for dimension $D=1$ and as open circles for $D=2$.
In both cases, we find strong differences between the static (Hartree-Fock) and the dynamical mean-field results. 
Hence, there are sizable effects resulting from local correlations. 
Comparing with the numerically exact DMRG data for the $D=1$ case, we furthermore see that the DMFT is not only able to qualitatively reproduce the crossover from the RKKY to the IIME regime but also predicts the magnetic properties of the system correctly on a quantitative level.

This is a remarkable finding as the $D=1$ case is actually the worst case from the DMFT perspective. 
One may exploit this to study more complex configurations of magnetic impurities in higher dimensions. 
Here, we will make use of the DMFT by accessing the single-electron excitation spectrum to study the fate of the flat band for a correlated system. 

\section{Coherent Green's function}

This can be done by looking at the low-frequency ``coherent'' part of the single-electron excitation spectrum. 
Within the DMFT framework, the coherent spectrum can be analyzed easily. 
The $\ff k$ and $\omega$ dependent single-electron $3\times 3$ Green's-function matrix is obtained from Dyson's equation as
\begin{eqnarray}
\label{Green}
  {\hat G}_{\ff k\sigma}(\omega)
  =
  \frac{1}{
  \omega + \mu - \hat \varepsilon - \hat t (\ff k) - {\hat \Sigma_\sigma}(\omega) 
  } 
  \, .
\label{eq:dyson}  
\end{eqnarray}
The $3\times 3$ hopping matrix $\hat t (\ff k)$ is given by Eq.\ (\ref{hopping}).
The ($\ff k$-independent) self-energy matrix ${\hat \Sigma_\sigma}(\omega) = \mbox{diag}(0,0,\Sigma_{\sigma}(\omega))$ and the matrix of on-site energies $\hat\varepsilon = \mbox{diag}(0,0,-U/2)$ are diagonal and non-zero on the impurity sites only.
Assuming that the system is a ferromagnetic Fermi-liquid, we can expand the self-energy for low frequencies: 
\begin{eqnarray}
\Sigma_{\sigma}(\omega) = a_\sigma + (1-z_\sigma^{-1}) \omega + {\cal O}(\omega^2) \, ,
\label{eq:expand}
\end{eqnarray}
where $a_\sigma$ is a real on-site energy shift and $0 \le z_\sigma \le 1$ is the quasi-particle weight. 
Inserting this into Eq.\ (\ref{Green}) and neglecting quasi-particle damping effects $\propto \omega^{2}$, we obtain the coherent part of the Green's function:
\begin{eqnarray}
{\hat G}^{\rm (con)}_{\ff k\sigma}(\omega) 
=
{\hat z}_\sigma^{1/2}
\frac{1}{
\omega + \mu - \hat \varepsilon_{\sigma}^{\rm (eff)} - {\hat t}_\sigma^{\rm (eff)}(\ff k)
}
{\hat z}_\sigma^{1/2} 
\, .
\label{eq:gcoh}
\end{eqnarray}
Comparing with the Green's function of the non-interacting system, we find (i) a spin-dependent shift of the effective on-site energy at the correlated site, $\hat \varepsilon_{\sigma}^{\rm (eff)} = \mbox{diag}[0,0,z_\sigma (a_{\sigma} - \frac{U}{2}) + \mu (1-z_\sigma)]$, (ii) a band renormalization
\begin{equation}
{\hat t}_\sigma^{\rm (eff)}(\ff k)
=
{\hat z}_\sigma^{1/2} \, {\hat t}(\ff k) \,  {\hat z}_\sigma^{1/2}
\label{eq:effhopp}
\end{equation}
with the quasi-particle weight matrix 
\begin{equation}
{\hat z}_\sigma
=
\left(
\begin{array}{ccc}
1 & 0 & 0 \\
 0 & 1 & 0 \\
 0& 0& z_\sigma
\end{array}
\right) \; , 
\end{equation}
as well as (iii) an overall scaling of the local quasi-particle density of states at the impurity sites.

\begin{figure}[t]
\centerline{\includegraphics[width=0.4\textwidth]{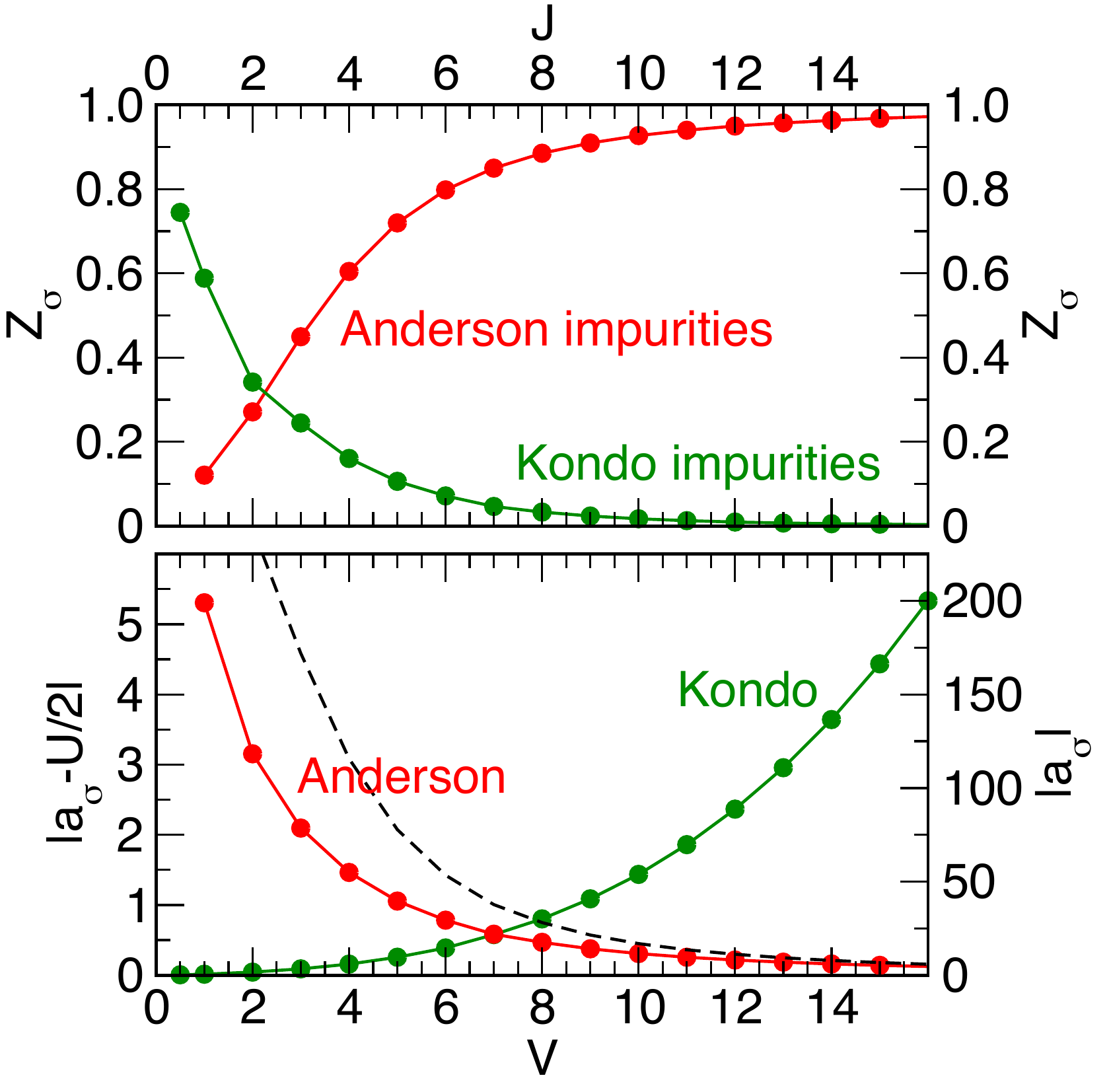}} 
\caption{
(Color online) Quasi-particle weight $z_{\sigma}$ and on-site energy shift $a_{\sigma}$ for a half-filled two-dimensional depleted Anderson lattice with $L = 1000 \times 1000$ sites for $U=16$ as functions of $V$ as obtained by DMFT (red circles, bottom and left axis) and $z_{\sigma} , a_{\sigma}$ for the case of Kondo impurities as a function of $J$ (green circles, top and right axis).
Spin dependence: $z_{\sigma} = z_{-\sigma}$, Anderson impurities: $a_{\downarrow} - U/2 = U/2 - a_{\uparrow} > 0$, Kondo impurities: $a_{\downarrow} = - a_{\uparrow} < 0$.
Dashed line: static mean-field result for $a_\sigma=U \langle n^{\rm (imp)}_{-\sigma} \rangle$ in the Anderson case.
Note that the DMFT self-energy is non-zero on the correlated sites in the Anderson model while for the Kondo model the self-energy is non-zero at the conduction-electron B sites.
}
\label{fig:qpw}
\end{figure}

The poles of the coherent Green's function close to $\omega =0$ determine the dispersion of the coherent quasi-particle band.
At half-filling, given for $\mu=0$, and in the spin-symmetric paramagnetic state with $a_{\sigma} = U/2$, the effective hopping matrix, Eq.\ (\ref{eq:effhopp}), is identical with the hopping matrix in the non-interacting case except for a renormalization $V \mapsto z_{\sigma}^{1/2} V$ of the hybridization. 
Hence, we find a flat quasi-particle band at $\omega = 0$. 
This is consistent with the expectation that a correlation-induced ``band narrowing'' of an already non-dispersive band does not have any effect.

However, this must be seen as an artifact of the DMFT as generally the self-energy and thus the parameters $z_{\sigma} = z_{\ff k\sigma}$ and $a_{\sigma} = a_{\ff k\sigma}$ acquire a $\ff k$-dependence which directly leads to a dispersion of the quasi-particle band.
Anyway, already on the DMFT level, the ferromagnetic long-range order implies that the coherent part of the excitation spectrum is dispersive since the spin-dependence of $a_{\sigma}$ also implies different effective on-site energies of the A and B sites of the bipartite lattice (see Sec.\ \ref{sec:non}).

Fig.\ \ref{fig:qpw} displays the parameters $z_{\sigma}$ and $a_{\sigma}$ as obtained from a DMFT calculation for a two-dimensional strongly correlated system ($U=2W=16$) as functions of $V$ (red symbols). 
A lattice with $L = 1000 \times 1000$ sites is sufficient to ensure that the results are not affected by finite-size effects.
In the ferromagnetic state at half-filling we have a spin-independent quasi-particle weight $z_{\sigma} = z_{-\sigma}$ but $a_{\uparrow} - U/2 < 0$ while $a_{\downarrow} - U/2 > 0$. 
The modulus of the deviation from $U/2$ (plotted in the figure) is spin-independent, similar but smaller than the corresponding static mean-field result $a_{\sigma} = U \langle n^{\rm (imp)}_{-\sigma} \rangle$. 

In the weak-$V$ limit, the system is effectively equivalent to the corresponding depleted Kondo lattice. 
Localization of electrons at the impurity sites and local-moment formation drive the system to a strongly correlated state with $z_{\sigma} \to 0$.
For large $V$, we find $z_{\sigma} \to 1$ and $a_{\sigma} \to U/2$: 
Due to strong charge fluctuations on the impurity sites, the system's low-frequency one-electron excitation spectrum is well described by non-interacting values of the parameters.

\section{Coherent density of states}
\label{sec:coh}

\begin{figure}[t]
\centerline{\includegraphics[width=0.4\textwidth]{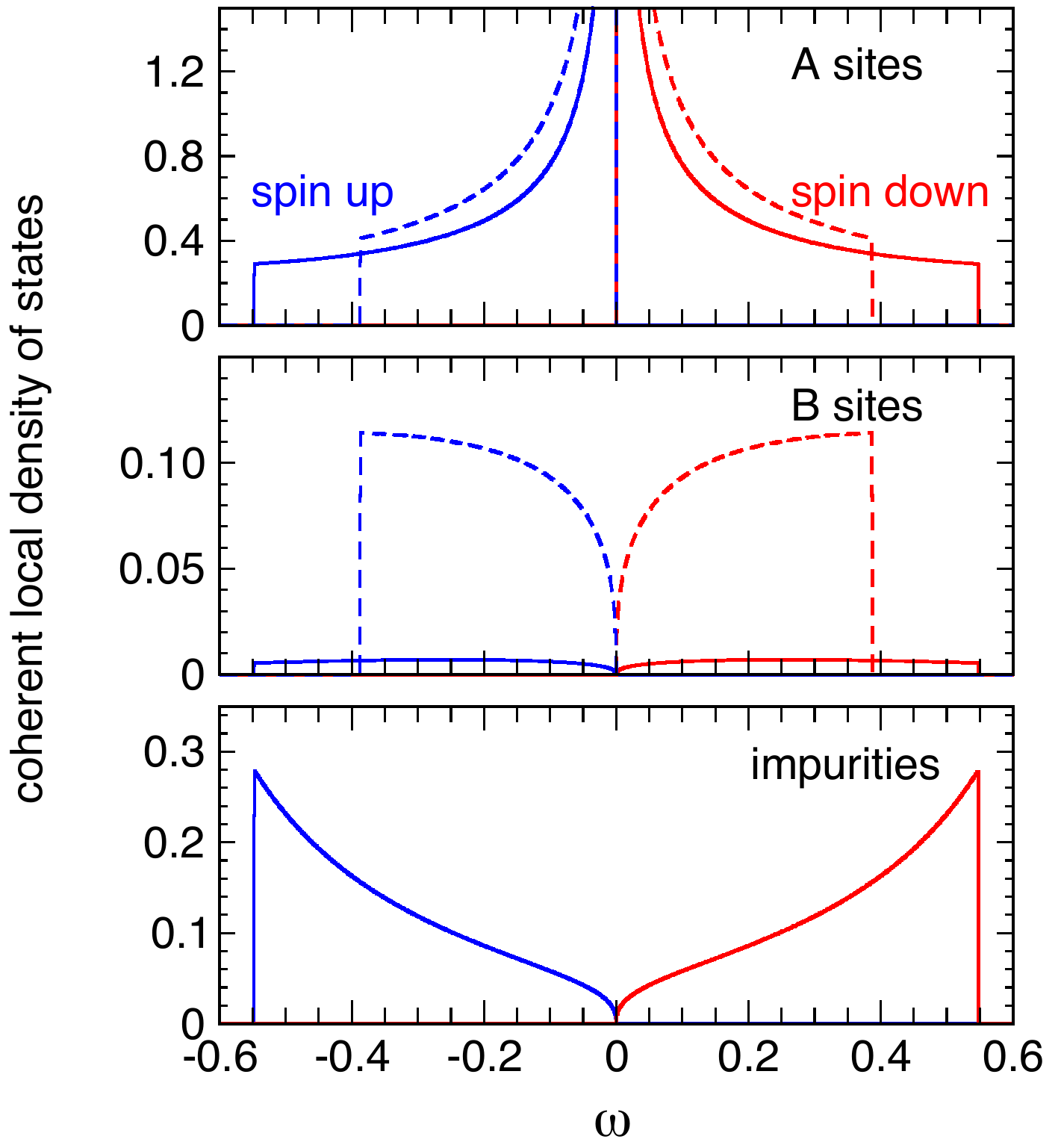}} 
\caption{
(Color online) Coherent part of the spin-dependent local density of states (solid lines) at the inequivalent sites $\alpha={\rm A,B,imp}$. 
Calculations for a two-dimensional depleted Anderson lattice with $L = 10000 \times 10000$ sites using the low-frequency parameters $z_{\sigma}, a_{\sigma}$ from a DMFT calculation for $U=16$ and $V=4$ (see Fig.\ \ref{fig:qpw}).
Blue: spin-up, red: spin-down coherent density of states.
Dashed lines: the same but for the case of Kondo impurities (DMFT calculation for $J=8$).
}
\label{fig:qps}
\end{figure}

Using the parameters from the DMFT calculation, we obtain the ``coherent'' part of the spin-dependent local density of states (DOS) projected onto A, B, and impurity sites,
\begin{equation}
  \rho^{\rm (coh.)}_{\alpha\sigma}(\omega) 
  = 
  - \frac{1}{\pi} \, \mbox{Im} \frac{2}{L} \sum_{\ff k} G_{\ff k\alpha\alpha\sigma}^{\rm (coh)}(\omega+i0^{+}) \: , 
\end{equation}  
see Fig.\ \ref{fig:qps}.
A somewhat larger lattice ($L=10000 \times 10000$) is necessary to completely suppress finite-size effects on the scale of the figure.

We discuss results for Hubbard interaction $U=16$. 
In this strong-coupling regime, the one-electron excitation spectrum shows incoherent Hubbard bands separated by an energy of the order of $U$.
The coherent part of the one-electron excitation spectrum consists of three quasi-particle bands which are separated by an energy of the order of $V$. 
Only the central band around the Fermi edge ($\omega = 0$) is plotted in Fig.\ \ref{fig:qps}, and here we focus on the low-frequency limit only where the Fermi-liquid form of the self-energy, Eq.\ (\ref{eq:expand}) is expected to hold.

First of all, the central quasi-particle band is dispersive, and the coherent part of the density of states has a finite width as discussed above.
However, the coherent DOS at the B sites is roughly two orders of magnitude smaller than at the A and at the impurity sites. 
This is still reminiscent of the flat-band picture that is found at $U=0$ or, at least within the DMFT, in the (metastable) paramagnetic phase where the central quasi-particle band is of A- and of the impurity-sites character only while the weight of the B sites is zero.

Interestingly, the coherent density of states exhibits a van Hove singularity at $\omega=0$ which may be seen as another reminiscence to the flat-band case at $U=0$ where the DOS includes a $\delta$-peak at $\omega=0$. 
In the correlated state, however, the singularity is weaker and only the DOS at the A sites is divergent.
Hence, the metallic character of the system is dominated by electron transport via the A sites. 

Furthermore, the coherent part of the DOS is fully polarized, i.e.\ $\uparrow$ and $\downarrow$ DOS do not overlap. 
Note that not only for the impurity and the A sites but also for B sites only the $\uparrow$ coherent DOS is occupied. 
However, $m_{\rm B} < 0$ as is seen from Fig.\ \ref{fig:mag} for $V/W=0.5$ and $U=16$ where the magnetic state of the system is found as intermediate between the RKKY and the IIME limits.
This implies that larger (negative) high-frequency contributions to the magnetic moment at the B sites must outweigh this effect.

Concluding, strong correlations drive the flat-band system to a ferromagnetic Fermi-liquid state with unconventional low-energy particle-hole excitations.
While the spin-$\uparrow$ as well as the spin-$\downarrow$ DOS are gapped, the total DOS is gapless.
Hence, low-energy particle-hole excitations contributing to the static conductivity must be accompanied by a spin flip.

\section{Kondo impurities}

This scenario can be tested in various ways: 
First, to simplify the system and to exclude the effect of charge fluctuations, we compare the results obtained for Anderson impurities with corresponding ones for Kondo impurities, i.e.\ we turn to the following Hamiltonian for a depleted Kondo lattice:
\begin{eqnarray}
{\cal H}
&=&
- t \sum_{\langle ij \rangle ,\sigma} ( a^\dagger_{i\sigma} b_{j\sigma} + \mbox{h.c.} )
+
\frac{J}{2}
\sum_{j \in B,\sigma\sigma'} 
{\ff S}_{k_{j}}  b^\dagger_{j\sigma} \hat {\ff \sigma}_{\sigma\sigma'} b_{j\sigma'}
\nonumber \\
&-& 
\mu \sum_{i \in A,\sigma} n_{i\sigma}^{(a)} 
-
\mu \sum_{j \in B,\sigma} n_{j\sigma}^{(b)} 
\, .
\label{eq:kondo} 
\end{eqnarray}
Here, $\ff S_{k_{j}}$ is a spin-$1/2$ at site $k_{j}$ that couples via an antiferromagnetic local exchange $J>0$ to the spins of conduction electrons at site $j$, the chemical potential is $\mu=0$ at half-filling, and $\hat {\ff \sigma}$ is the vector of Pauli matrices. 

For the DMFT calculations this depleted Kondo-lattice model is self-consistently mapped onto an effective impurity model where the correlated impurity consists of a B-site $j$ with the local spin at $k_{j}$ attached and is embedded in an uncorrelated bath. \cite{NMK00,OKK09a}
An effective impurity model with $n_{\rm s} = 9$ sites (plus the local spin $1/2$) is considered in practice and treated with the Lanczos technique.

Dyson's equation is again given by Eq.\ (\ref{eq:dyson}) but now the hopping and the self-energy in $\ff k$ space are $2\times 2$ matrices:
\begin{equation}
  \hat t ({\bf k}) 
  = 
  \left(
  \begin{array}{cc}
  0 & \varepsilon({\bf k}) \\
  \varepsilon({\bf k}) & 0 \\
  \end{array}
  \right) 
  \; , \qquad
  \hat \Sigma_{\sigma}(\omega)
  =
  \left(
  \begin{array}{cc}
  0 & 0 \\
  0 & \Sigma_{\sigma}(\omega) \\
  \end{array}
  \right) 
  \: .
\label{2x2}
\end{equation}
Expanding the self-energy for $\omega\to 0$, Eq.\ (\ref{eq:expand}), leads to Eq.\ (\ref{eq:gcoh}) with an effective on-site energy at the correlated B site, $\hat \varepsilon_{\sigma}^{\rm (eff)} = \mbox{diag}[0,z_\sigma a_{\sigma} + \mu (1-z_\sigma)]$, and to a band renormalization,
${\hat t}_\sigma^{\rm (eff)}(\ff k) = {\hat z}_\sigma^{1/2} \, {\hat t}(\ff k) \,  {\hat z}_\sigma^{1/2}$,
with the quasi-particle weight matrix 
\begin{equation}
{\hat z}_\sigma
=
\left(
\begin{array}{ccc}
1 & 0 \\
0 & z_\sigma
\end{array}
\right) \; .
\end{equation}
In the metastable paramagnet at half-filling, the effective hopping matrix is given by the non-interacting one, except for a renormalization $\varepsilon(\ff k) \to z^{1/2} \varepsilon(\ff k)$ of the bare dispersion. 
This leads to a completely flat quasi-particle band.
As in the Anderson case, however, the ferromagnetic long-range order makes the band dispersive since $a_{\sigma} \ne 0$ implies different effective on-site energies of the A and B sites.

The parameters $z_{\sigma}$ and $a_{\sigma}$ characterizing the low-frequency part of the self-energy are shown in Fig.\ \ref{fig:qpw} as a function of $J$. 
The quasi-particle weight $z_{\sigma} = z_{-\sigma}$ at the B sites and the modulus of the on-site energy shift $|a_{\sigma}| = |a_{-\sigma}|$ are spin-independent but $a_{\uparrow} < 0$ while $a_{\downarrow} > 0$. 

When comparing the low-energy electronic structure of the Anderson and the Kondo model with each other, it is important to recall that the DMFT construction is different:
While for the Anderson case a correlated site with $U>0$ in the lattice model must be identified with the impurity site of the effective Anderson impurity model, the spin $1/2$ {\em and} the attached B site have to be considered as the impurity in the Kondo case. \cite{NMK00,OKK09a}
Consequently, the self-energies are non-zero on correlated ($U>0$) sites in the Anderson and non-zero on B-sites in the Kondo case.

In the weak-$J$ limit, the Anderson and the Kondo model can be mapped onto each other. 
Here, electrons are only weakly scattered from the impurities and thus $z_{\sigma} \to 1$.
For strong $J$, on the other hand, an impurity spin and an electron at a B site form an almost local Kondo singlet. 
Scattering of A-site electrons from these local Kondo singlets tends to localize electrons at A sites and results in a strongly correlated Fermi-liquid state with small $z_{\sigma}$. 
Furthermore, the IIME mechanism drives the system to the ferromagnetic state as discussed in Sec.\ \ref{sec:iime}.

The resulting ``coherent'' part of the spin-dependent local density of states (DOS) projected onto A and B sites is shown in Fig.\ \ref{fig:qps} for $J=8$ (dashed lines).
The DOS resulting from the coherent low-energy band around the Fermi edge has a finite width. 
Its projection onto the B sites is an order of magnitude smaller than the projection on the A sites.
The coherent DOS is again fully polarized and also exhibits the same (singular) structure at $\omega=0$ that was found for the Anderson case.
We conclude that even in the strong-$J$ limit where the Kondo lattice does not map onto the Anderson lattice, the low-frequency physics is qualitatively the same.

\section{$D=1$ DMFT Green's function}

\begin{figure}[t]
\centerline{\includegraphics[width=0.4\textwidth]{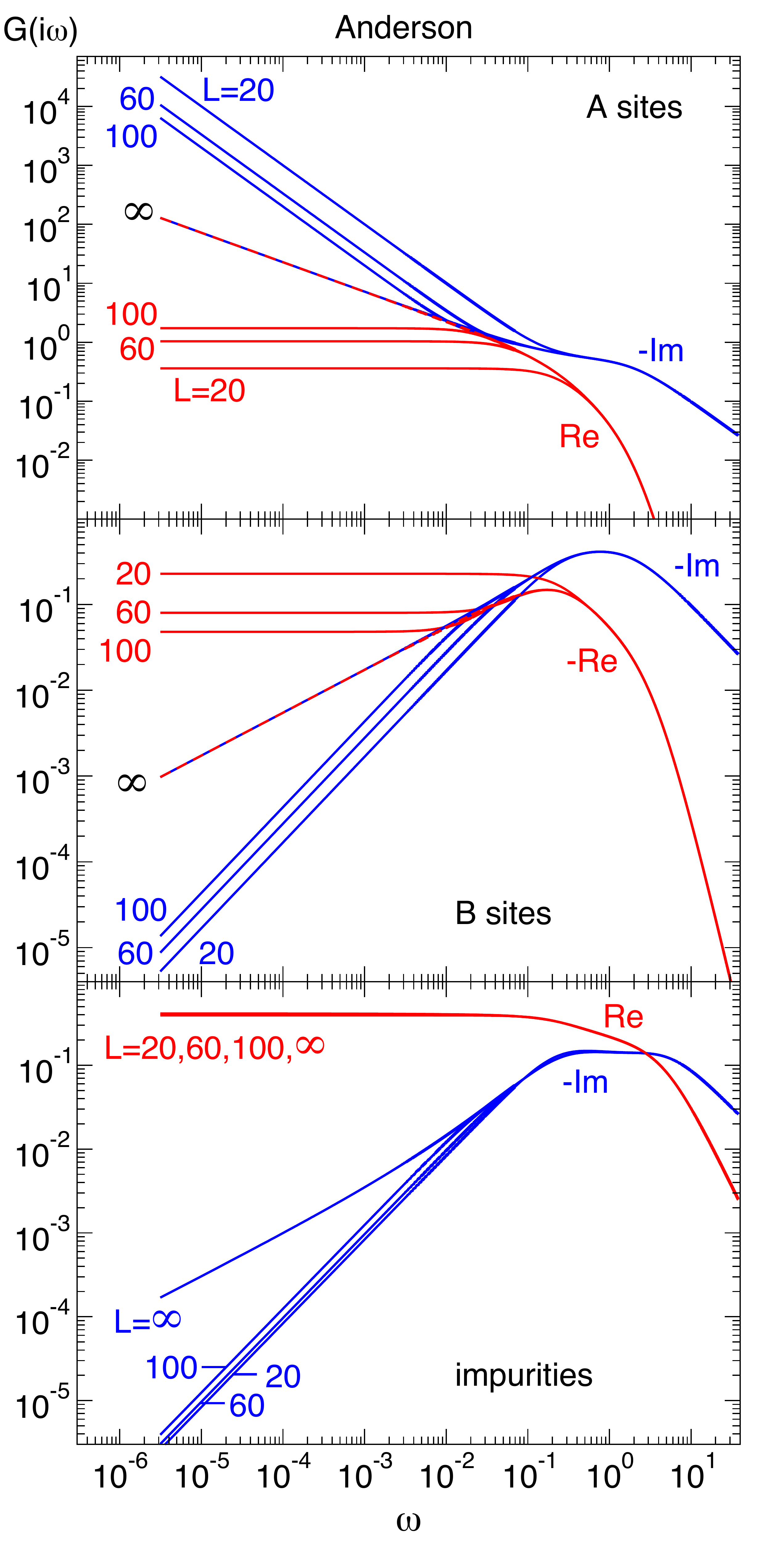}}
\caption{(Color online) 
Spin-$\uparrow$ Green's function for the one-dimensional depleted Anderson model as obtained from DMFT for $V=1$ and $U=8$.
Imaginary (blue) part, $- \mbox{Im} \, G_{\alpha\alpha,\sigma=\uparrow}(i\omega)$, 
and real part (red), $\pm \mbox{Re}\, G_{\alpha\alpha,\sigma=\uparrow}(i\omega)$, for $\alpha={\rm A,B,imp}$ and different system sizes $L$ as indicated.
Note the double logarithmic scale.
}
\label{fig:gand}
\end{figure}

\begin{figure}[t]
\centerline{\includegraphics[width=.4\textwidth]{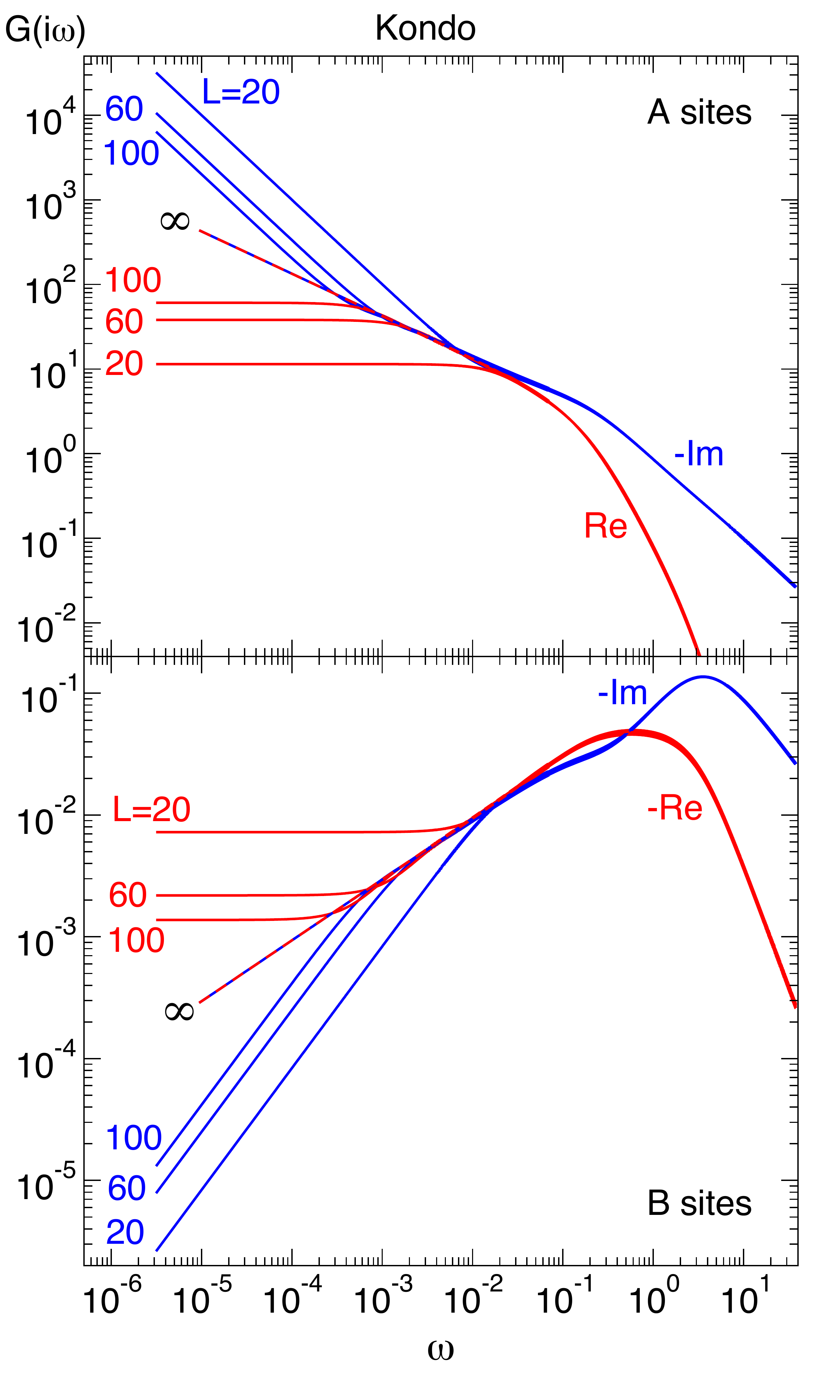}}
\caption{(Color online) 
The same as in Fig.\ \ref{fig:gand} but for the depleted Kondo lattice at $J=5$.
}
\label{fig:gkon}
\end{figure}

As the DMFT is able to quantitatively describe the correlated magnetic ground state even for the one-dimensional system, it is worthwhile to discuss the $D=1$ single-particle excitation spectra. 
The local spin-$\uparrow$ Green's functions for A, B and impurity sites in the range of small imaginary frequencies are shown in Figs.\ \ref{fig:gand} and \ref{fig:gkon} for the case of Anderson ($V=1, U=8$) and for the case of Kondo impurities ($J=5$), respectively. 
The spin-$\downarrow$ Green's functions are fixed by the relations
\begin{eqnarray}
  \mbox{Re} \, G_{\alpha\alpha,\downarrow}(i\omega) 
  &=& 
  - \mbox{Re} \, G_{\alpha\alpha,\uparrow}(i\omega) 
  \; ,
  \nonumber \\
  \mbox{Im} \, G_{\alpha\alpha,\downarrow}(i\omega) 
  &=& 
  \mbox{Im} \, G_{\alpha\alpha,\uparrow}(i\omega)
\label{eq:phsym}
\end{eqnarray}
which are enforced by particle-hole symmetry. 

Results are displayed for different finite systems with $L=20,60,100$ and for the infinite system. 
As can be seen from the figures, finite-size effects start to play a dominant role in the low-frequency regime for $\omega \lesssim 1/L$.
The low-frequency behavior of the Green's function for $L=\infty$ can be obtained from the Dyson equation (\ref{eq:dyson}) by computing the matrix inverse analytically and by replacing the $\ff k$-sum with an integration $\int d\varepsilon$ weighted with the non-interacting density of states $\rho(\varepsilon)$. 
For $D=1$ we have $\rho(\varepsilon) = 1/(\pi \sqrt{4-\varepsilon^{2}})$, and the $\varepsilon$-integration can be done analytically. 
For low frequencies, i.e.\ using Eq.\ (\ref{eq:expand}), we then find for the Anderson case:
\begin{eqnarray}
  G_{{\rm AA},\uparrow}(i\omega) &=& \frac{V}{2t} \frac{1}{\sqrt{U/2-a_\uparrow}} \frac{1}{\sqrt{i\omega}}
\: , 
\nonumber \\ 
  G_{{\rm BB},\uparrow}(i\omega) &=& - \frac{1}{2Vt} \sqrt{U/2 - a_\uparrow} \sqrt{i\omega }
\: , 
\nonumber \\
  G_{{\rm imp},\uparrow}(i\omega) &=& \frac{1}{U/2 - a_\uparrow} - \frac{V}{2t} \frac{\sqrt{i\omega}}{\left(U/2 - a_\uparrow \right)^{3/2}} 
\: , 
\label{eq:lowand}
\end{eqnarray}
while for the case of Kondo impurities
\begin{eqnarray}
  G_{{\rm AA},\uparrow}(i\omega) &=& \frac{1}{2t} \frac{\sqrt{a_\uparrow}} {\sqrt{i\omega}} 
  \; ,
\nonumber \\ 
  G_{{\rm BB},\uparrow}(i\omega) &=& -\frac{1}{2t}\frac{\sqrt{i \omega}}{\sqrt{a_\uparrow}} 
   \: .
\label{eq:lowkon}
\end{eqnarray}
This perfectly agrees with the data shown in Figs.\ \ref{fig:gand} and \ref{fig:gkon}. 
As in the $D=2$ case (see Fig.\ \ref{fig:qps}), we find a diverging A-site quasi-particle density of states at the Fermi edge $\propto - \mbox{Im} \, G_{AA,\sigma}(i\omega)$ for $\omega \to 0$ while the B-site and the impurity DOS vanish. 
In fact, for $D=2$ the low-frequency quasi-particle DOS is similar,
$\rho^{\rm (coh.)}_{A\uparrow}(\omega) \propto \sqrt{\omega^{-1} \ln |\omega|}$ 
and 
$\rho^{\rm (coh.)}_{B\uparrow}(\omega) \propto \sqrt{\omega \ln |\omega|}$, but with additional logarithmic corrections that are traced back to the van Hove singularity of the $D=2$ non-interacting density of states.

Concluding, the interaction-induced renormalization of the low-energy one-particle excitation spectrum generates dispersive quasi-particle bands with characteristic van Hove singularities.
Let us emphasize that this is the mean-field and Fermi-liquid picture for the excitation spectrum which can be provided by the DMFT but which is likely to be invalidated for $D=1$ below some low-frequency scale by coupling to bosonic long-wave-length modes.

For the finite-size systems, the DMFT picture is expected to be more adequate as those modes are cut by the finite-size gap. 
Figs.\ \ref{fig:gand} and \ref{fig:gkon} show that, below a certain frequency scale of the order of $1/L$, the A-site Green's function behaves as $G_{{\rm AA},\uparrow}(\omega) = \mbox{const.} \times \omega^{-1}$ while for the B-site Green's function (and likewise for the impurity Green's function) we have $G_{{\rm BB},\uparrow}(\omega) = - \mbox{const.} \times \omega$ with positive constants. 
According to the Lehmann representation of the Green's function for a system of finite size,
\begin{equation}
  G_{\alpha\beta,\sigma}(\omega) = \sum_{n} \frac{z_{\alpha\beta, n}}{\omega - \omega_{n}} \; , 
\end{equation}
this implies that there is a pole at zero frequency, $\omega_{0} = 0$, with a finite weight $z_{{\rm AA}, 0} > 0$ on the A sites while $z_{{\rm BB},0} = z_{{\rm imp},0}=0$.
Generally, $z_{\alpha\alpha,0} = g^{-1} \sum_{m,n} |\langle m | c^{\dagger}_{\alpha \uparrow}| n \rangle|^{2}$, where $m,n$ label the mutually orthogonal ground states of the system, and $g$ is the ground-state degeneracy. 
Here, a two-fold degeneracy of the ground state arises from the fact that, for any finite $L$, two eigenvalues of the effective hopping matrix $\hat \varepsilon_{\sigma}^{\rm (eff)} + {\hat t}_\sigma^{\rm (eff)}(\ff k)$ in Eq.\ (\ref{eq:gcoh}) are vanishing, namely at $k=\pi/2$ and at $k=-\pi/2$.
The corresponding eigenvectors have 100\% A-character. 
The same behavior of $G_{{\rm AA},\uparrow}(\omega)$ is also found for higher dimensions but finite $L$, where there is at least a two-fold ground-state degeneracy. 

\section{Charge susceptibility}

A diverging total density of states at the Fermi edge implies a diverging charge response to a change of the chemical potential, i.e.\ a diverging charge susceptibility $\kappa = \partial n / \partial \mu$ where $n$ is the average total particle number per site.
Hence, we can check the DMFT results by comparing with DMRG data for $\kappa$ available for $D=1$, finite $L$ and half-filling.
$\kappa$ is related to the charge gap,
\begin{equation}
   \Delta_{\rm c}
   =
   \frac{E_{0}(L+2,M_{\rm tot}) + E_{0}(L-2,M_{\rm tot}) - 2 E_{0}(L,M_{\rm tot})}{4}
   \: ,
\end{equation}
which is obtained from the ground-state energy $E_{0}(N,M_{\rm tot})$ in the sector with total particle number $N$ and $z$-component of the total spin $M_{\rm tot}$.
We have
\begin{equation}
  \kappa 
  = 
  \lim_{L \to \infty}
  \frac{1}{L \Delta_{c}} \: .
\end{equation}
Calculations have been performed for sectors with different $M_{\rm tot}$.
A finite $\kappa$ is only obtained if $M_{\rm tot} \ne \pm S_{\rm tot}$, and the results for different $M_{\rm tot}$ agree. 
For $M_{\rm tot} = S_{\rm tot}$, corresponding to the symmetry-broken state that is realized in the related DMFT calculation, however, $\kappa$ vanishes in the limit $L\to \infty$. 
The reason is that particle-hole excitations with arbitrarily low excitation energy would have to be accompanied by a spin flip, as it has already been discussed in Sec.\ \ref{sec:coh} on the DMFT level, while $\kappa$ is sensitive to the spin-independent charge response only.

\begin{figure}[t]
\centerline{\includegraphics[width=0.4\textwidth]{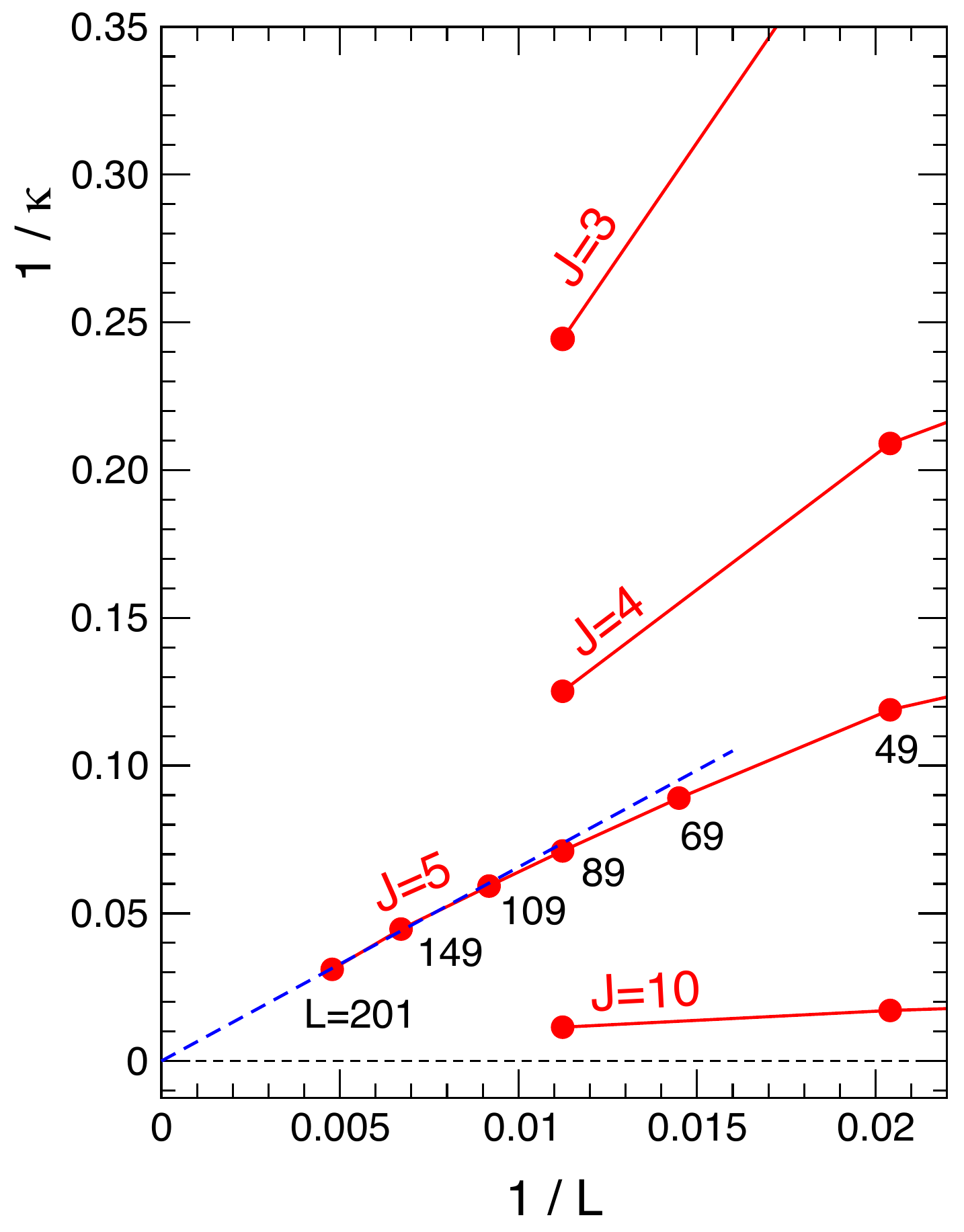}  }
\caption{
(Color online) Inverse charge susceptibility$1/\kappa$ as a function of inverse system size $L$ for the one-dimensional depleted Kondo lattice. Result obtained by DMRG for different $J$ as indicated and for $M_{\rm tot}=0$. The dashed line is a fit to the data for $J=5$.
}
\label{fig:kappa}
\end{figure}

Fig.\ \ref{fig:kappa} displays DMRG data for $\kappa$ obtained for the $M_{\rm tot}=0$ sector of the Kondo lattice at different $J$ in the strong-coupling regime and different system sizes $L$. 
As is demonstrated by finite-size scaling for $J=5$, the charge susceptibility diverges with increasing system size. 
This is fully consistent with the interpretation of the DMFT results. 
The fact that $\kappa$ is increasing with increasing $J$ can be attributed to the increasing tendency to  form local Kondo singlets at the B sites and thereby to localize electrons at A sites. 

A diverging $\kappa$ in many cases indicates an instability of the ground state towards phase separation (see Ref.\ \onlinecite{EKPV07} for an example of phase separation in a different fermion-lattice model at half-filling):
An S-shaped trend of $n(\mu)$ with $\kappa \to \infty$ for $\mu \to \mu_{1,2}$ implies that there is a finite range of chemical potentials, $\mu_{1} < \mu < \mu_{2}$, where $\partial n / \partial \mu < 0$, i.e.\ a range where a homogeneous phase is thermodynamically unstable and where an inhomogeneous state composed of macroscopically separated regions with different fillings has a lower grand potential.
In the present case, however, the divergence of $\kappa$ originates from the van Hove singularity of the density of states at the Fermi edge and will thus be absent for fillings off half-filling.
We therefore expect that there is no {\em finite} $\mu$ range, i.e.\ $\mu_{1}=\mu_{2}$, or, stated differently, that $\mu(n)$ has a saddle point at half-filling rather than a maximum and a minimum at fillings below and above half-filling, respectively.
This is corroborated by the DMRG calculations for $N=L\pm 2$ where we did not find any tendencies for the formation of an inhomogeneous ground state.

\section{Conclusions}

The periodic Anderson model with regularly depleted impurities, or the depleted Kondo-lattice model, provide the most simple setup to study ferromagnetic order of localized moments induced by the indirect RKKY magnetic exchange.
On a bipartite lattice with conduction-electron hopping between nearest-neighbor sites and for a half-filled conduction band, the period of the effective RKKY exchange couplings is commensurate with the lattice and is ferromagnetic if the distance between the impurities is $d=2$.

We have studied the magnetic properties of this model on a one-dimensional chain and on the two-dimensional square lattice in the RKKY regime, i.e.\ at weak hybridization strength $V$, or weak Kondo coupling $J$, but also the crossover to the strong-$V$ or strong-$J$ regime.
Besides the RKKY approach, there are several and rather different theoretical concepts that apply to this case.

A variant of the Lieb-Mattis theorem for the Kondo model \cite{LM62a,She96} predicts that there is a ferromagnetic ground state with a macroscopically large total spin quantum number $S_{\rm tot}$ among the different ground states (if there is ground-state degeneracy). 
In certain geometries, $S_{\rm tot}$ is less by one as compared to the naive application of the RKKY theory, and the (finite-size) Kondo effect must be considered in addition for an explanation. \cite{SGP12}

Perturbative approaches in the weak-coupling \cite{SGP12} and the strong-coupling (strong $V$ or $J$) cases \cite{STP13} can be employed to clarify whether the ground state is degenerate or not (apart from the trivial $2S_{\rm tot}+1$ spin degeneracy). 
For strong $V$ or $J$ in particular, the concept of the inverse indirect magnetic exchange (IIME) explains why there is ferromagnetic order despite the fact that the impurity magnetic moments are Kondo screened, namely local magnetic moments at the intermediate A sites are formed and couple magnetically via virtual excitations of the Kondo singlets.
For the one-dimensional case the crossover from the RKKY to the IIME limit is nicely seen in the local ordered magnetic moment and in spin correlation functions with the help of density-matrix renormalization. 

Another route to ferromagnetic order that may apply to the class of systems considered here is the concept of flat-band ferromagnetism. \cite{Mie91a,Tas92,Tas97} 
In fact, the bipartite geometry, the restriction to nearest-neighbor hopping and the depletion of impurities with $d=2$ straightforwardly implies the emergence of a flat band at the Fermi energy of the non-interacting ($U=0$) system.
This implies that the $U=0$ ground-state energy is highly degenerate and that the ``fully polarized'' state where the flat band is exactly half-filled with spin-$\uparrow$ electrons only is among the ground states.
It is remarkable that the computed $V$- and site-dependent local magnetic polarizations in this state already roughly capture the main trend, namely ferromagnetic order sustained by the impurities with $m_{\rm imp} \to 1$ but $m_{\rm A} \to 0$ in the RKKY limit $V \to 0$ while for the IIME limit $V\to \infty$ we have $m_{\rm imp} \to 0$ but $m_{\rm A} \to 1$ and ferromagnetism is sustained by conduction electron localized at A sites.

The Slater determinant with a fully polarized flat band would be the only ground state for any finite $U$ if the concept of flat-band ferromagnetism applies.
However, already Hartree-Fock theory shows that this is not the case.
Furthermore, comparing the predictions of Hartree-Fock theory with the essentially exact DMRG data, we conclude that the ground state is highly correlated rather than a simple Fermi sea for intermediate or strong $U$. 

Hartree-Fock theory in fact provides a fairly good but still rough description of the symmetry broken state. 
More surprising is the fact that the dynamical mean-field theory, even for the one-dimensional case, yields {\em quantitatively} almost exact results as is verified by comparing with the DMRG data.
We conclude that the depletion of the impurities drives the system to a state which is strongly correlated (and thus different from the static mean-field state) but where the correlations are mainly temporal rather then spatial (and thus accessible to the dynamical mean-field approach).
Technically, the finite distance $d=2$ between the impurities implies that the non-local contributions to the electron self-energy become negligibly small as
already the lowest-order non-local corrections scale with the third power of the fourth-nearest neighbor element of the non-interacting Green's function.
This offers the exciting perspective that a comparatively simple DMFT approach can be employed to quantitatively describe the indirect magnetic coupling and the resulting magnetic order of nanostructures in higher spatial dimensions, e.g.\ on a two-dimensional metallic surface.

Future work will have to address the magnetic and correlated electronic structure of depleted Anderson- and Kondo-lattice models away from the particle-hole symmetric point off half-filling.
This brings in different new aspects, such as, for example, the absence of a Lieb-Mattis theorem or the incommensurability of the RKKY couplings with the lattice constant. 
A rather complex magnetic phase diagram can be expected.
Furthermore, it will be interesting to study the filling dependence of the single-particle excitation spectrum.

Here, for the case of half-filling, we have found a rather unconventional low-frequency electronic structure using the DMFT, namely a gapless metallic spectrum but with a completely filled $\uparrow$ and empty $\downarrow$ ``coherent'' part of interacting density of states.
Ferromagnetic order at $U>0$ has been seen to necessarily result in a finite quasi-particle dispersion of the originally (for $U=0$) entirely flat band, i.e.\ a ``correlated flat band'' is no longer flat. 
Due to the bipartite lattice structure and the manifest particle-hole symmetry at half-filling the interacting density of states develops (weaker) singularities at the Fermi edge which can be understood as van Hove singularities of the low-frequency coherent quasi-particle band structure.
These result in a diverging homogeneous charge susceptibility as could also be verified for the one-dimensional case by DMRG but are not expected to drive the system to a phase-separated state. 

\acknowledgments

Financial support of this work by the Deutsche Forschungsgemeinschaft within the SFB 668 (project A14) and within the excellence cluster ``The Hamburg Centre for Ultrafast Imaging - Structure, Dynamics and Control of Matter at the Atomic Scale'' is gratefully acknowledged.


\appendix

\section{First-order perturbation theory in $U$}
\label{appendix:1OC}

According to Stoner's criterion and due to the occurrence of a flat band at the Fermi energy, the weakly interacting depleted periodic Anderson model at half-filling should be unstable towards ferromagnetic ordering. 
Here, we summarize the results of first-order perturbation theory in $U$ for the Hamiltonian Eq.\ (\ref{eq:ham}).

The mean-field decoupling Eq.\ (\ref{MF_decoupling}) is exact up to first order in $U$ and provides us with a simplified Hamiltonian
${\cal H_{\rm MF}}$ which is given by $\ca H_{0}$, Eq.\ (\ref{H0}), with $\hat t(\ff k)$ replaced by $\hat t(\ff k) + \hat {\Sigma}_{\sigma}^{(1)}$, and where
$\hat {\Sigma}_{\sigma}^{(1)}$ is a $3\times 3$ matrix with ${\Sigma}_{\alpha,\alpha', \sigma}^{(1)} = 0$ for all $\alpha,\alpha' = \mbox{A, B, imp}$ except for $\alpha=\alpha'=\mbox{imp}$ where we have ${\Sigma}_{{\rm imp,imp} ,\sigma}^{(1)} \equiv {\Sigma}_{{\rm imp} ,\sigma}^{(1)} = U (n_{\rm imp,-\sigma} - 1/2)$ with $n_{{\rm imp},\sigma} = \langle n_{k_{j} \sigma}^{(c)} \rangle$. 
Obviously, there is no difference to the non-interacting case for the paramagnetic state.

For the ferromagnetic state, $\hat t(\ff k) + \hat {\Sigma}_{\sigma}^{(1)}$ is diagonalized by the unitary transformation 
\begin{eqnarray}
\label{Appendix:Q_sigma}
\hat Q_{\sigma}^{(1)}({\bf k}) = \hat Q_{0}({\bf k})
\left(1+ {\Sigma}_{{\rm imp} ,\sigma}^{(1)} \hat \Delta({\bf k})\right) \, 
\end{eqnarray}
up to first order in $U$ where $\hat Q_{0}({\bf k})$ is given by Eq.\ (\ref{eq:qmat}) and
\begin{eqnarray}
\label{Appendix:U1}
\hat \Delta({\bf k})=\frac{1}{\sqrt{2}\, \xi({\bf k})^3}\left(
\begin{array}{ccc}
 0 &V \varepsilon({\bf k}) & -\frac{1}{2\sqrt{2}}V^2  \\
-V \varepsilon({\bf k}) & 0 & V \varepsilon({\bf k})\\
\frac{1}{2\sqrt{2}}V^2 & -V \varepsilon({\bf k}) & 0
\end{array}
\right)
\, .
\nonumber \\
\end{eqnarray}
Therewith, up to first order in $U$, we find the eigenvalues
\begin{eqnarray}
\eta_{1,\sigma}^{(1)}({\bf k})&=&-\xi({\bf k})-\mu + \frac{V^2}{2\xi({\bf k})^2} {\Sigma}_{{\rm imp} ,\sigma}^{(1)} \; ,\nonumber \\
\eta_{2,\sigma}^{(1)}({\bf k})&=&-\mu + \frac{\varepsilon_{\bf {\bf k}}^2}{\xi({\bf k})^2} {\Sigma}_{{\rm imp} ,\sigma}^{(1)} \; ,\nonumber\\
\eta_{3,\sigma}^{(1)}({\bf k})&=&\xi({\bf k})-\mu + \frac{V^2}{2\xi({\bf k})^2} {\Sigma}_{{\rm imp} ,\sigma}^{(1)}\; , 
\end{eqnarray}
[compare with Eq.\ (\ref{eq:spectrum})] where $\xi({\bf k})=\sqrt{\varepsilon({\bf k})^2+V^2}$.
Analogously to Eqs.\ (\ref{eq:magdef}) and (\ref{eq:ns}), we get the site-dependent magnetic polarizations
\begin{eqnarray}
m_{\rm A}^{(1)} 
&=&
m_{\rm A}^{(0)} -\frac{2}{L}\sum_{\bf k} \frac{3V^2\varepsilon({\bf k})^2}{4\xi({\bf k})^5} U m_{\rm imp}^{(0)} 
\; , \nonumber \\
m_{\rm B}^{(1)}
&=&
-\frac{2}{L}\sum_{\bf k}\frac{V^2}{4\xi({\bf k})^3} U  m_{\rm imp}^{(0)} 
\; , \nonumber \\
m_{\rm imp}^{(1)}
&=&
m_{\rm imp}^{(0)}+\frac{2}{L}\sum_{\bf k}\frac{4 V^2 \xi({\bf k})^2-3 V^4}{4\xi({\bf k})^5} U m_{\rm imp}^{(0)} 
\; .
\nonumber \\
\end{eqnarray}
The non-interacting values are given by Eq.\ (\ref{eq:ms}).
Note that $m_{\rm A} + m_{\rm B} + m_{\rm imp} =1$.
The first-order effect of the Hubbard-type interaction is thus to increase $m_{\rm imp}$ and $|m_{\rm B}|$ and to decrease $m_{\rm A}$.

The total energy of magnetic ground state, 
\begin{eqnarray}
\label{Appendix:GS}
E_{\rm F}
=
-\frac{4}{L} \sum_{\bf k} \xi({\bf k})-\frac{U}{4}(n_{\rm imp}^2+m_{\rm imp}^2)
\end{eqnarray}
($n_{\rm imp} = n_{\rm imp,\uparrow} + n_{\rm imp,\downarrow}$) is lower than the energy of the paramagnetic state which is obtained by setting $m_{\rm imp}=0$ and which is equal to the total energy of the non-interacting system.

\end{document}